\def\boldxi{\mbox{\boldmath$\xi$}}

\documentclass[onecolumn]{emulateapj}
\usepackage{color}

\begin{document}

\slugcomment{\apj, submitted 7 Jul 2016, accepted 27 Jun 2017 (Printed \today)}
\title{A Global Model For Circumgalactic and Cluster-Core Precipitation}
\author{G. Mark Voit\altaffilmark{1,2}, 
	    Greg Meece\altaffilmark{1},
	    Yuan Li\altaffilmark{3},
	    Brian W. O'Shea\altaffilmark{1,4}, 
	    Greg L. Bryan\altaffilmark{5},
	    Megan Donahue\altaffilmark{1} 
	    }  
\altaffiltext{1}{Department of Physics and Astronomy,
                     Michigan State University,
                     East Lansing, MI 48824} 
\altaffiltext{2}{voit@pa.msu.edu}
\altaffiltext{3}{Department of Astronomy,
                     University of Michigan,
                     Ann Arbor, MI} 
\altaffiltext{4}{Department of Computational Mathematics, Science, and Engineering,
                     Michigan State University,
                     East Lansing, MI 48824} 
\altaffiltext{5}{Department of Astronomy,
                     Columbia University,
                     New York, NY} 

\begin{abstract}
We provide an analytic framework for interpreting observations of multiphase circumgalactic gas that is heavily informed by recent numerical simulations of thermal instability and precipitation in cool-core galaxy clusters.   We start by considering the local conditions required for the formation of multiphase gas via two different modes: (1) uplift of ambient gas by galactic outflows, and (2) condensation in a stratified stationary medium in which thermal balance is explicitly maintained.  Analytic exploration of these two modes provides insights into the relationships between the local ratio of the cooling and freefall time scales (i.e., $t_{\rm cool} / t_{\rm ff}$), the large-scale gradient of specific entropy, and development of precipitation and multiphase media in circumgalactic gas.  We then use these analytic findings to interpret recent simulations of circumgalactic gas in which global thermal balance is maintained.  We show that long-lasting configurations of gas with $5 \lesssim \min (t_{\rm cool} / t_{\rm ff}) \lesssim 20$ and radial entropy profiles similar to observations of local cool-core galaxy cluster cores are a natural outcome of precipitation-regulated feedback.  We conclude with some observational predictions that follow from these models.   This work focuses primarily on precipitation and AGN feedback in galaxy cluster cores, because that is where the observations of multiphase gas around galaxies are most complete.  However, many of the physical principles that govern condensation in those environments apply to circumgalactic gas around galaxies of all masses.  
\end{abstract}

\vspace*{2.0em}

\section{Introduction}

\setcounter{footnote}{0}

The relationship between thermal instability and galaxy formation is a classic topic in theoretical astrophysics that has recently come back into fashion.  Its reemergence has been driven by the need to understand how accretion onto supermassive black holes regulates cooling and star formation in galaxy-cluster cores.  Both observational and theoretical evidence is accumulating in support of the idea that development of a multiphase medium through condensation in the vicinity of a supermassive black hole triggers strong feedback that limits further condensation \citep[e.g.,][]{ps05,Soker06,Cavagnolo+08,ps10,McCourt+2012MNRAS.419.3319M,Sharma_2012MNRAS.420.3174S,Gaspari+2012ApJ...746...94G,Gaspari+2013MNRAS.432.3401G,Gaspari_2015A&A...579A..62G,Voit_2015Natur.519..203V,Li_2015ApJ...811...73L,Tremblay2016}.  Perhaps most intriguingly, if there is a similar link between feedback heating and condensation of circumgalactic gas around smaller galaxies, then this regulation mechanism has much broader implications for galaxy evolution \citep[e.g.,][]{Soker10_galform,Sharma+2012MNRAS.427.1219S,Voit_PrecipReg_2015ApJ...808L..30V}.

\subsection{Heritage of the Topic}

All modern discussions of thermal instability in the context of galaxy formation are rooted in the classic work of \citet{ReesOstriker1977MNRAS.179..541R}, \citet{Binney1977ApJ...215..483B}, and \citet{Silk1977ApJ...211..638S}, which themselves owe a debt to \citet{Hoyle_1953ApJ...118..513H}.  These landmark papers derived the maximum stellar mass of a galaxy ($\sim 10^{12} \, M_\odot$) by comparing the time for gas to fall through a galaxy's potential well to the time required to cool from the potential's virial temperature.  If the cooling time is less than the freefall time, then infalling gas can potentially condense and fragment into star-forming clouds via thermal instability.  That can happen relatively easily in galaxy-scale objects with virial temperatures $\lesssim 10^7$~K but is more difficult to arrange in hotter, more massive sytems, leading to a natural division between the mass scales of individual galaxies and those of galaxy groups and clusters.

Many subsequent papers have made interesting use of the cooling-time to freefall time ratio to analyze how galaxies form and evolve \citep[e.g.,][]{BFPR_1984Natur.311..517B,FallRees1985ApJ...298...18F,MallerBullock_2004MNRAS.355..694M}.  However, some form of negative feedback is necessary to explain the inefficient transformation of a galaxy's gas supply into stars \citep[e.g.,][]{Larson_1974MNRAS.169..229L,wr78,DekelSilk1986ApJ...303...39D,WhiteFrenk1991ApJ...379...52W,Baugh_1998ApJ...498..504B,SomervillePrimack_1999MNRAS.310.1087S,KauffmannHaehnelt_2000MNRAS.311..576K}.  A complete understanding of the relationship between thermal instability and galaxy formation must therefore account for interplay between radiative cooling and the energetic feedback that opposes cooling.

\citet{Field65} worked out many of the fundamental features of astrophysical thermal instability but did not consider the complications that arise when thermal instability couples with buoyancy.  The most comprehensive analytical treatment of that coupling is by \citet{bs89}, who astutely summarized much of the preceding work.  Balbus \& Soker were primarily concerned with the development of inhomogeneity in galaxy-cluster cores.  At the time, hot gas in the cores of many galaxy clusters was suspected to condense at rates $\sim 10^{2-3} \, M_\odot \, {\rm yr}^{-1}$ \citep[e.g.,][]{Fabian94}, but models of homogeneous cooling flows into the cluster's central galaxy produced X-ray surface-brightness profiles with central peaks far greater than were observed.  This conundrum led to speculation that the mass inflow rate could decline inward because of thermal instability and spatially distributed condensation, which would reduce the radiative losses required to maintain a steady state at the center of the flow \citep[e.g.,][]{Thomas_1987MNRAS.228..973T,WhiteSarazin_analytical_1987ApJ...318..612W,WhiteSarazin_data_1987ApJ...318..621W,WhiteSarazin_numerical_1987ApJ...318..629W}.  Upon closer examination, this speculation was found to be problematic, because buoyancy generally tends to suppress the development of thermal instability \citep[e.g.,][]{Cowie_1980MNRAS.191..399C,Nulsen_1986MNRAS.221..377N,bs89}.  The attention of the field therefore gradually shifted away from steady-state cooling-flow models in favor of models in which feedback from a central active galactic nucleus compensates for cooling \citep[e.g.,][]{TaborBinney1993MNRAS.263..323T,BinneyTabor_1995MNRAS.276..663B,Soker+01,mn07,McNamaraNulsen2012NJPh...14e5023M}

\subsection{Renaissance of the Topic}

More than two decades later, the coupling between buoyancy and thermal instability is being re-examined, because the presence of inhomogenous gas in galaxy-cluster cores consisting of multiple phases that are orders of magnitude cooler and denser than the ambient medium now appears closely linked with the ratio of cooling time to freefall time \citep[e.g.,][]{McCourt+2012MNRAS.419.3319M,Gaspari+2012ApJ...746...94G,VoitDonahue2015ApJ...799L...1V,Voit_2015Natur.519..203V}.  In those studies, the cooling time is typically defined with respect to the specific heat at constant volume, so that $t_{\rm cool} \equiv [ 3kT / n_e \Lambda (T) ] (n/2n_i)$, where $\Lambda (T)$ is the usual cooling function at temperature $T$, and the number densities of electrons, ions, and gas particles are $n_e$, $n_i$, and $n$, respectively.  The freefall time $t_{\rm ff} \equiv (2r/g)^{1/2}$ is defined with respect to the local gravitational potential $g$ at radius $r$.  Given these definitions, a floor appears to be present near $t_{\rm cool} / t_{\rm ff} \approx 10$ in the radial cooling-time profiles of the ambient hot gas in galaxy clusters.  A large majority of the cluster cores known to contain multiphase gas have minimum values of $t_{\rm cool} / t_{\rm ff}$ within a factor of 2 of this floor.  Conversely, almost all the clusters without multiphase gas have $\min (t_{\rm cool} / t_{\rm ff}) > 20$.

\citet{McCourt+2012MNRAS.419.3319M} and \citet{Sharma_2012MNRAS.420.3174S} interpreted this relationship between multiphase gas and $t_{\rm cool} / t_{\rm ff}$ as resulting from amplification of initially small perturbations by thermal instability.  Under conditions of global thermal balance,  numerical simulations of thermal instability in a plane-parallel potential by \citet{McCourt+2012MNRAS.419.3319M} showed that $t_{\rm cool} / t_{\rm ff} \lesssim 1$ was required for thermal instability resulting in condensation, but the simulations of \citet{Sharma_2012MNRAS.420.3174S} in a spherically symmetric potential indicated that condensation could happen in spherical systems with $t_{\rm cool} / t_{\rm ff} \lesssim 10$.  The threshold value of $t_{\rm cool} / t_{\rm ff}$ was therefore assumed to be geometry dependent. Subsequent simulations implementing more sophisticated treatments of feedback appeared to corroborate that assumption because they showed that condensation-fueled accretion of cold gas onto a central black hole can lead to long-lasting self regulation with $\min (t_{\rm cool} / t_{\rm ff}) \approx 10$ \citep[e.g.,][]{Gaspari+2012ApJ...746...94G,LiBryan2014ApJ...789...54L,Li_2015ApJ...811...73L,Prasad_2015ApJ...811..108P}.

In the meantime, it has become clear that the critical ratio of $t_{\rm cool} / t_{\rm ff}$ is not a geometry-dependent manifestation of local thermal instability. For example, numerical simulations by \citet{Meece_2015ApJ...808...43M} of thermal instability in a plane-parallel potential under conditions seemingly quite similar to those adopted by \citet{McCourt+2012MNRAS.419.3319M} showed that condensation could occur at the midplane of systems with any value of $t_{\rm cool} / t_{\rm ff}$.  Also, \citet{ChoudhurySharma_2016MNRAS.457.2554C} have presented a detailed thermal stability analysis of systems in global thermal balance showing that the growth rates of linear perturbations are largely independent of the gravitational potential's geometry.  So why, then, do both real and simulated galaxy-cluster cores appear to self-regulate at $\min (t_{\rm cool} / t_{\rm ff}) \approx 10$?

\subsection{Precipitation-Regulated Feedback}

The most general answer seems to involve a phenomenon that we have come to call precipitation.  As feedback acts on a galaxy-scale system, the outflows it drives can promote condensation of the hot ambient medium by raising some of it to greater altitudes \citep[e.g.,][]{Revaz_2008A&A...477L..33R,LiBryan2014ApJ...789..153L,McNamara_2016arXiv160404629M}.  Adiabatic uplift promotes condensation by lowering the $t_{\rm cool} / t_{\rm ff}$ ratio of the uplifted gas (see \S \ref{sec-tc_isoK}). This process is loosely analogous to the production of raindrops during adiabatic cooling of uplifted humid gas in a thunderstorm---hence, the name ``precipitation.''  As in a thunderstorm, the condensates rain down toward the bottom of the potential well after they form.  In simulations, this rain of cold gas into the galaxy at first provides additional fuel for feedback and temporarily boosts the strength of the outflows, but eventually those strengthening outflows add enough heat to the ambient medium to raise $t_{\rm cool} / t_{\rm ff}$ high enough to stop the condensation.  Precipitation is therefore naturally self-regulating.

A prescient series of ``cold feedback'' papers by Soker \& Pizzolato anticipated many features of the precipitation mechanism now seen in simulations \citep{ps05,Soker06,Soker08,ps10}.  They proposed a feedback cycle in cluster cores in which energetic AGN outbursts produce a wealth of non-linear density perturbations, the densest of which cool, condense, fall back toward the black hole, and provide more fuel for accretion.  In this scenario, the cooling times of the blobs must be short enough that $t_{\rm cool} \lesssim t_{\rm ff}$ and also shorter than the time interval between AGN heating outbursts.  They argued that this source of accretion fuel could potentially provide much more fuel than Bondi accretion from the hot medium alone while responding to changes in the ambient cooling time far more quickly.  They also noted that a shallow central entropy gradient would promote condensation.

Shortly thereafter, Gaspari and collaborators produced sets of numerical simulations in which such a cold feedback loop was realized \citep[][but see also Sharma et al. 2012b]{Gaspari+2012ApJ...746...94G,Gaspari+2013MNRAS.432.3401G,Gaspari_2015A&A...579A..62G}.  In these simulations, cold clouds form through thermal instability and accrete toward the center of the simulation volume while the heating mechanism needed to maintain approximate global thermal balance stimulates turbulence.  This turbulence is critical, because it ensures a steady supply of cold gas blobs with low specific angular momentum, which can plunge to the center through a process the authors call ``chaotic cold accretion.''

The proliferation of terminology has a way of making these mechanisms seem more different than they really are.  Precipitation-regulated feedback, as described in this paper, is a ``cold feedback'' mechanism that fuels a central black hole through ``chaotic cold accretion.''  Here we are proceeding with the term ``precipitation,'' despite the prior existence of these other terms, because the processes that promote thermal instability and condensation in circumgalactic gas may be responsible for more than just the feeding of black holes.  Therefore, they merit a more general term.

\subsection{Implications for Galaxy Evolution}

Precipitation is potentially of broader interest because it may link a galaxy's time-averaged star-formation rate with the multiphase structure of its circumgalactic gas.  Observations of circumgalactic absorption lines are showing that the masses of gas and metals within a few hundred kpc of a galaxy are at least as great as the galaxy's mass in the form of stars \citep[e.g.,][]{Tumlinson_2011Sci...334..948T}.  Also, the amount of circumgalactic gas at intermediate temperatures ($10^5$-$10^6$~K) appears closely linked with a galaxy's star-formation rate.  Observations of the circumgalactic medium at other wavelengths likewise show rich multiphase structure \citep[e.g.,][]{Putman_2012ARA&A..50..491P}, which has been challenging for simulations to reproduce \citep[e.g.,][]{Hummels_2013MNRAS.430.1548H,Ford_2016MNRAS.459.1745F}.

\citet{Sharma+2012MNRAS.427.1219S} proposed that thermal instability, through precipitation-regulated feedback, places a lower limit of $t_{\rm cool} / t_{\rm ff} \approx 10$ on the ambient density of circumgalactic gas \citep[but see also][]{Meece_2015ApJ...808...43M}.  \citet{Voit_PrecipReg_2015ApJ...808L..30V} built on that idea to show how precipitation-regulated feedback could be responsible for governing not only the relationships between stellar mass, metallicity, and stellar baryon fraction observed among galaxies but also the relationship between a galaxy's stellar velocity dispersion and the mass of its central black hole.  However, they did so without having a satisfactory explanation for the crucial assumption that $\min(t_{\rm cool} / t_{\rm ff}) \approx 10$ or a complete model for the global structure of precipitation-regulated systems.

\subsection{Purpose of the Paper}

This paper's purpose is to propose a global context for interpreting observations of multiphase gas around galaxies, as well as the numerical simulations that strive to reproduce those observations, in terms of the long legacy of theoretical papers on astrophysical thermal instability. Many of the theoretical results derived here were published decades ago by others, most notably by \citet{Defouw_1970ApJ...160..659D,Cowie_1980MNRAS.191..399C,Nulsen_1986MNRAS.221..377N,Malagoli_1987ApJ...319..632M,Loewenstein_1989MNRAS.238...15L}; and \citet{bs89}.  Our re-derivations of them are intended to provide a common conceptual framework for a meta-analysis of simulations in which precipitation occurs.   We are focusing on simulations of precipitation in galaxy-cluster cores, because that is where the observations of multiphase gas around galaxies and interactions of outflows with the circumgalactic medium are most complete.  However, many of the physical principles that govern condensation in those environments apply to circumgalactic gas around galaxies of all masses.

\begin{figure*}[t]
\begin{center}
\includegraphics[width=6.75in, trim = 0.0in 0.0in 0.0in 0.0in]{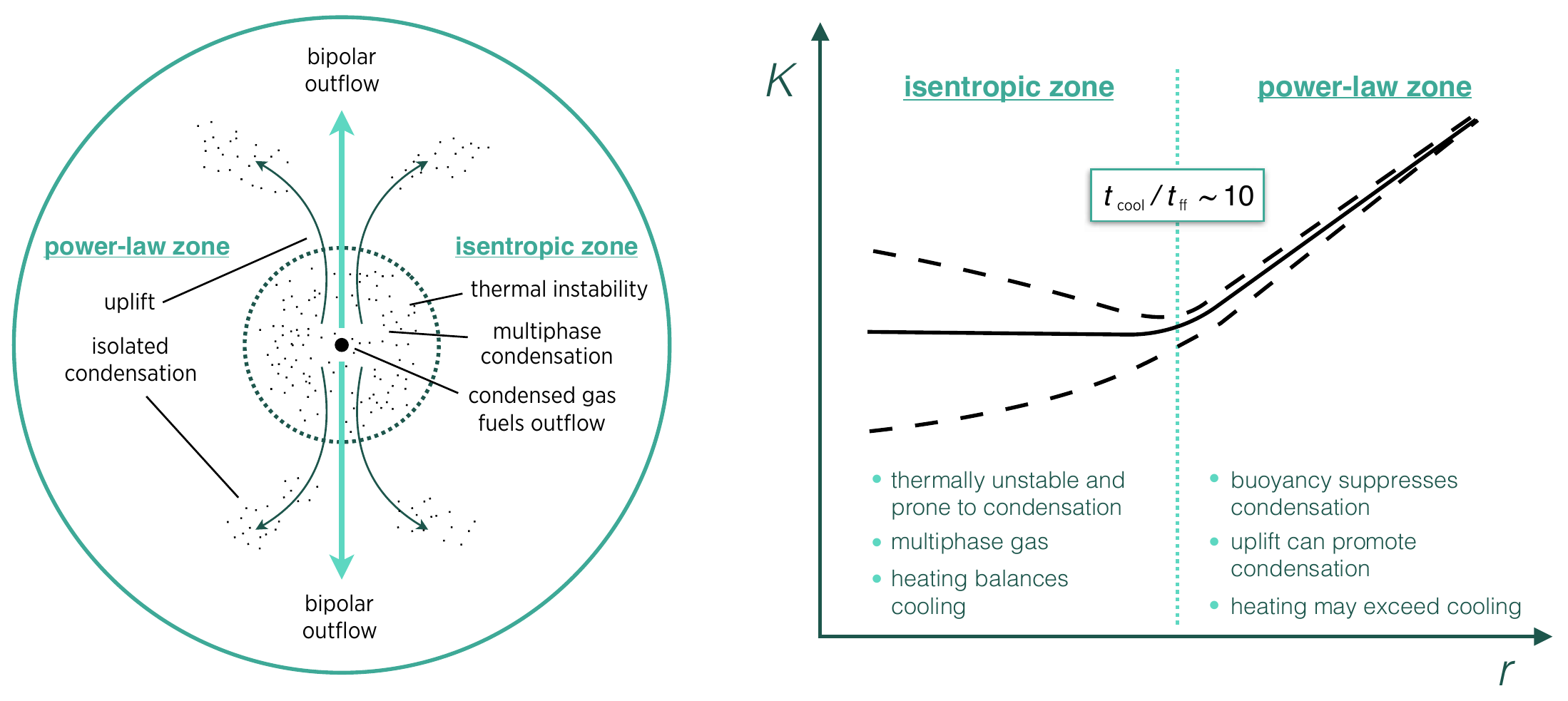} \\
\end{center}
\caption{ \footnotesize 
This schematic cartoon outlines the main ideas presented in the paper.  On the left is a diagram of a galactic environment in which feedback is active.  Accretion of condensed gas onto the central black hole releases feedback energy, and a bipolar outflow distributes that energy over a large volume.  In order for the system to develop a well-regulated feedback loop, it must separate into two zones, an inner ``isentropic zone'' and an outer ``power-law zone'' in which the specific entropy ($K \equiv kTn_e^{-2/3}$) follows $d \ln K / d \ln r  \approx 1$, as observed in both real and simulated galaxy-cluster cores  (\S \S \ref{sec-GlobalBalance},\ref{sec-Illustrations}).  The power-law entropy gradient allows buoyancy to limit the growth of thermal instability 
(\S \S\ref{sec-GeneralConsiderations},\ref{sec-TI_NumericalSimulations}), implying that condensation in the power-law zone requires uplift of lower-entropy gas (\S \ref{sec-tc_isoK}).  In contrast, buoyancy cannot suppress thermal instability in the isentropic zone, which proceeds to develop multiphase structure, as indicated by the dashed lines showing the dispersion in $K$ at each $r$ (\S \S \ref{sec-TI_NumericalSimulations},\ref{sec-Illustrations}).  Phenomenologically, the ratio of cooling time to free-fall time in the ambient medium is observed to reach a minimum value in the range $5 \lesssim t_{\rm cool} / t_{\rm ff} \lesssim 20$ at the boundary between these zones in cluster cores with multiphase gas (\S \ref{sec-Illustrations}).  Such a system cannot remain in a steady feedback-regulated state with a central cooling time $\lesssim 1$~Gyr unless a large proportion of the feedback energy is thermalized {\em outside} of the central isentropic region (\S\ref{sec-GlobalBalance}).
\vspace*{1em}
\label{fig-CartoonVersion}}
\end{figure*}

\subsection{A Readers' Guide}
\label{sec-ReadersGuide}

Busy readers may wish to be selective in deciding which sections of this long paper will reward their close attention.  For them, we have prepared this guide, along with a cartoon (Figure \ref{fig-CartoonVersion}) that sketches out the main ideas.

\begin{itemize}

\item  The next three sections (\S \S \ref{sec-tc_isoK}-\ref{sec-TI_NumericalSimulations}) consider the local conditions required for condensation and precipitation.  Two different modes of precipitation emerge from those considerations.  One is analogous to rain that is stimulated by uplift of humid gas in Earth's atmosphere, because of the role that adiabatic cooling plays in bringing on condensation.  The other is analogous to drizzle or fog, in that condensation happens without uplift, when the conditions are right. 

\begin{itemize}

\item Section \ref{sec-tc_isoK} initiates the discussion with a brief reminder about the deep connections between adiabatic uplift and condensation.  It also presents a short calculation showing that ambient gas uplifted at speeds comparable to a halo's circular velocity is likely to condense if it initially has $t_{\rm cool} / t_{\rm ff} \lesssim 10$.  This finding suggests that the ambient medium around a galaxy cannot persist in a state with $t_{\rm cool} / t_{\rm ff} \ll 10$ if there is significant vertical circulation.  Galactic outflows in which the energy source is fueled by condensation therefore tend to drive the ambient medium toward $t_{\rm cool} / t_{\rm ff} \approx 10$.

\item Section \ref{sec-GeneralConsiderations} outlines the general conditions necessary for thermal instability without uplift to progress to condensation in a hydrostatic medium that is thermally balanced within each equipotential layer.  The main result is that the condition for condensation to occur depends not only on $t_{\rm cool} / t_{\rm ff}$ but also on the slope of the entropy gradient:  Thermal instability leads to condensation only if a low-entropy perturbation can cool faster than it sinks to a layer of equivalent entropy.  This is not a new result, but its significance is often not fully appreciated.

\item Section \ref{sec-TI_NumericalSimulations} uses the results of \S \ref{sec-GeneralConsiderations} to interpret recent simulations of thermal instability in circumgalactic gas.  In particular, it calls attention to the critical role of the global entropy gradient in determining where condensation can occur and where it is suppressed by buoyancy.  The main result is that media in thermal balance are prone to condensation in regions where the large-scale entropy gradient is flat.  However, buoyancy tends to delay the onset of condensation if $t_{\rm cool} / t_{\rm ff} \gg 1$.

\end{itemize}

\item The following two sections (\S \S \ref{sec-GlobalBalance},\ref{sec-Illustrations}) apply the findings from the first part of the paper to interpret the global evolution of simulated galactic systems in which condensation fuels feedback.  Our objective is to understand why those systems end up in long-lasting configurations with $5 \lesssim \min (t_{\rm cool} / t_{\rm ff}) \lesssim 20$ and with radial entropy profiles in agreement with observations of multiphase galaxy-cluster cores.

\begin{itemize}

\item Section \ref{sec-GlobalBalance} examines recent simulations of condensation in globally balanced but locally unstable galactic systems in light of the findings summarized in \S \ref{sec-TI_NumericalSimulations}.  It points out that feedback is {\em required} for the development of a multiphase medium, because phase separation cannot happen in a globally balanced medium without a flow of free energy through the system.  Condensation also cannot happen through linear thermal instability in regions with a significant entropy gradient and $t_{\rm cool} \gg t_{\rm ff}$.  Steady self-regulation of a precipitating system therefore favors a global configuration with a shallow inner entropy gradient and a steeper outer entropy gradient (Figure \ref{fig-CartoonVersion}).  The shallow inner gradient promotes the precipitation needed for fuel, while the steeper outer entropy gradient prevents condensation from running away into a cooling catastrophe.  The boundary between these regions tends to be where $t_{\rm cool} / t_{\rm ff}$ reaches a minimum value $\sim 10$, for reasons outlined in \S \ref{sec-tc_isoK}.  In order to ensure long-term global stability with a central cooling time $\lesssim 1$~Gyr, much of the feedback energy must propagate beyond the isentropic zone before thermalizing.  This finding has deep implications for implementations of black-hole feedback in numerical simulations.

\item Section~\ref{sec-Illustrations} puts all these pieces together to interpret the time-dependent behavior of precipitation-regulated feedback in a numerical simulation from \citet{Li_2015ApJ...811...73L}.  It shows that unopposed cooling leads to a power-law entropy gradient at the center of the system, which focuses condensation onto the central black hole.  The feedback response disrupts that central gradient out to where $t_{\rm cool} / t_{\rm ff} \approx 10$ in the ambient medium.  Much of the gas uplifted from that central region can be induced to condense, particularly if it is inhomogeneous.  The system then settles into a long-lasting steady state in which condensed gas fuels the outflow and feedback maintains the isentropic central region at a level corresponding to $5 \lesssim \min (t_{\rm cool} / t_{\rm ff}) \lesssim 20$.  Catastrophic cooling is prevented because much of the outflow's energy is thermalized outside of the isentropic zone.  When the condensed gas is depleted, the outflow shuts down.  Cooling then proceeds almost homogeneously, because there is no source of free energy to promote phase separation.  However, thermal instability eventually initiates condensation near the outer edge of the isentropic region, at the minimum of $t_{\rm cool} / t_{\rm ff}$ in the ambient medium.  Newly condensed gas subsequently reignites feedback, and the cycle repeats.

\end{itemize}

\end{itemize}

\newpage 

\noindent 
The paper concludes with two sections that acknowledge the loose ends (\S \ref{sec-LooseEnds}) and present some concluding thoughts about how the overall model can be tested (\S \ref{sec-Summary}), followed by an appendix that constructs a useful toy model for global configuration changes of precipitation-regulated systems.

\section{Uplift and Condensation}
\label{sec-tc_isoK}

In the precipitation framework for self-regulating feedback, adiabatic uplift is one of the main mechanisms for causing condensation of circumgalactic gas. Transporting gas upward without changing its entropy increases its propensity to condense by lowering its $t_{\rm cool} / t_{\rm ff}$ ratio.  Most of the change in this critical ratio comes from the increase in $t_{\rm ff}$ owing to uplift, because the change in cooling time produced by adiabatic uplift is relatively modest.

\subsection{Adiabatic Uplift}

To illustrate this point, consider a blob of gas that begins at radius $r_1$ with entropy $K_1$ in an isothermal potential with constant circular velocity $v_c$.  The freefall time in this potential is linearly proportional to radius, and the cooling time of the blob depends on its pressure.  When written in terms of entropy $K$ and gas pressure $P$, this cooling time is
\begin{eqnarray}
  t_{\rm cool} & \; = \; & \frac {3 K^{6/5}} {P^{1/5} \, \Lambda[T(P,K)]}  
  			\left( \frac {n} {n_e} \right)^{1/5} 
			\left( \frac {n} {2n_i} \right) 
			\\
           & \; \approx \; & 235 \, {\rm Myr} 
                         \left( \frac {K} {10 \, {\rm keV \, cm^2}} \right)^{6/5}
                         \left( \frac {P} {k \cdot 10^6 \, {\rm K \, cm^{-3}}} \right)^{-1/5}
                         \left( \frac {\Lambda} {2 \times 10^{-23} \, {\rm erg \, cm^3 \, s^{-1}}} \right)^{-1}
			\; \; .
\end{eqnarray}
For gas with the temperatures and abundances typical of galaxy cluster cores and large elliptical galaxies ($10^7 \, {\rm K} \lesssim T \lesssim 3 \times 10^7 \, {\rm K}$), the cooling function $\Lambda (T)$ is nearly independent of temperature, meaning that $t_{\rm cool}$ changes very little as gas adiabatically expands or contracts.  One can account for the temperature dependence of $\Lambda$ by defining $\lambda \equiv d \ln \Lambda / d \ln T$ and noting that $T \propto K^{3/5} P^{2/5}$, to obtain $t_{\rm cool} \propto K^{(6 - 3 \lambda)/5} P^{-(1 + 2 \lambda)/5}$.  Adiabatic uplift of a gas blob in a stratified atmosphere therefore changes the ratio of cooling time to free fall time within the blob according to
\begin{equation}
  \left( \frac {t_{\rm cool}} {t_{\rm ff}}  \right)_r =  \left( \frac {t_{\rm cool}} {t_{\rm ff}} \right)_{r_1}
							  \left( \frac {r_1} {r} \right)^{1 - 2 \beta (1+2\lambda)/5} \; \; ,
\end{equation}
where $d \ln P / d \ln r = - 2 \beta$ with $\beta \equiv \mu m_p v_c^2 / 2 kT$ in a hydrostatic ambient medium.  Uplift by an order of magnitude in radius consequently lowers the $t_{\rm cool}/t_{\rm ff}$ ratio local to the blob by roughly an order of magnitude, given the typical $\beta$ and $\lambda$ values of circumgalactic gas ($0.5 \lesssim \beta \lesssim 1$ and $-1 \lesssim \lambda \lesssim 0.5$), as long as magnetic fields sufficiently suppress thermal conduction.

This relationship helps to explain why precipitation-driven feedback tends to push a galaxy's volume-filling ambient medium into a state with $\min( t_{\rm cool} / t_{\rm ff} ) \approx 10$.  If the central gas is so dense that $t_{\rm cool} / t_{\rm ff}  \ll 10$, then outflows propelled by feedback can promote additional condensation of the ambient medium with relatively modest amounts of adiabatic uplift.  Newly condensed clouds that have not achieved escape velocity can then rain back down into the galaxy to fuel more star formation and AGN feedback, and the rain will continue until dissipation of feedback energy raises the ambient value of $\min( t_{\rm cool} / t_{\rm ff} )$ enough to inhibit condensation of uplifted gas.  Then feedback finally succeeds in stopping the accumulation of cold clouds within the galaxy.  The critical value of $\min( t_{\rm cool} / t_{\rm ff} )$ at which cessation of condensation should happen is hard to calculate from first principles, but simulations of the process indicate that it happens around $\min( t_{\rm cool} / t_{\rm ff} ) \approx 10$ \citep[e.g.,][]{Gaspari+2012ApJ...746...94G,Li_2015ApJ...811...73L,Prasad_2015ApJ...811..108P,Meece_2016arXiv160303674M}.

\subsection{Ballistic Condensation}
\label{sec-ballistic}

A more quantitative relationship between uplift and condensation can be derived from the cooling history of an uplifted gas blob that follows a purely radial ballistic trajectory without incurring any heat input.  For simplicity, consider a gravitational potential in which $v_c$ is constant with radius, so that $\phi (r) = v_c^2 \ln (r/r_{\rm max})$, where $r_{\rm max}$ is the radius at which the blob reaches its apex. This implies
\begin{equation}
  r = r_{\rm max} \exp \left( - \frac {v^2} {2 v_c^2} \right) \; \; ,
\end{equation}
where $v$ is the radial velocity, which leads to the equation of motion
\begin{equation}
  \ddot{r} = \dot{v} = - \frac {v_c^2} {r_{\rm max}}  \exp \left( \frac {v^2} {2 v_c^2} \right) \; \; .
\end{equation}
We are interested in knowing the blob's time of flight $\Delta t$ from an initial velocity $v_1$ to a final velocity $v_2$, which can be determined by integrating the equation of motion:
\begin{equation}
  \Delta t \; = \; \frac {r_{\rm max}} {v_c^2} \int_{v_2}^{v_1} \exp \left( \frac {v^2} {2 v_c^2} \right) dv 
              \; = \; \frac {\pi^{1/2}} {2} 
                     \left[ {\rm erf} \left( \frac {v_1} {\sqrt{2} v_c}  \right) 
                      - {\rm erf} \left( \frac {v_2} {\sqrt{2} v_c}  \right) \right]
                      t_{\rm ff} (r_{\rm max}) 
                      \; \; .
\end{equation}
The maximum time of flight is then $\max (\Delta t) = \pi^{1/2} t_{\rm ff} (r_{\rm max})$, and setting $v_1 = - v_2 = v_c$ gives $\Delta t \approx 1.2 \, t_{\rm ff} (r_{\rm max})$.

\begin{figure*}[t]
\begin{center}
\includegraphics[width=6in, trim = 0.0in 0.1in 0.0in 0.0in]{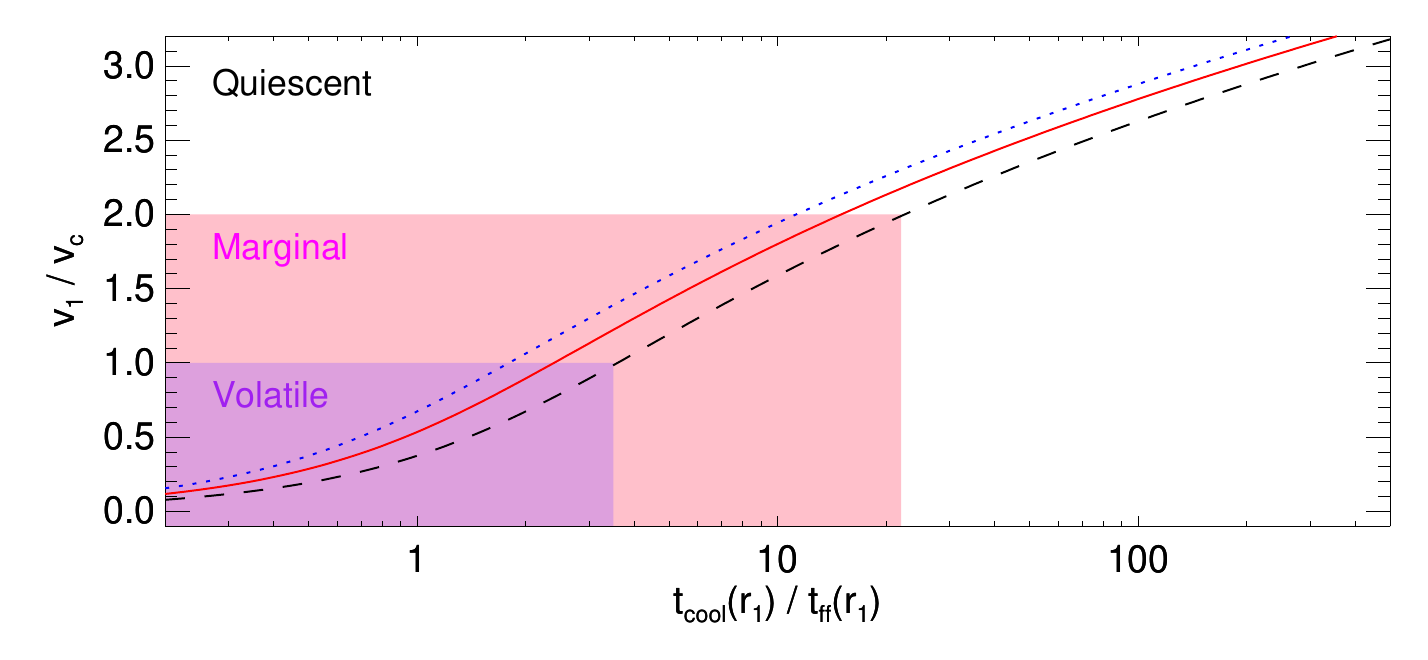} \\
\end{center}
\caption{ \footnotesize 
Values of uplift velocity $v_1$ required for a gas blob to condense on a ballistic trajectory starting at radius $r_1$ as a function of its value of $t_{\rm cool} / t_{\rm ff}$ at $r_1$.  Lines correspond to different values of the cooling-function slope $\lambda \equiv d \ln \Lambda / d \ln T$, with a dotted blue line showing $\lambda = 1/2$ (characteristic of free-free cooling at $T \gtrsim 10^7$~K), a dashed black line showing $\lambda = -1$ (characteristic of collisional emission-line cooling at $10^5 \, {\rm K} < T < 10^7 \, {\rm K}$, and a solid red line showing $\lambda = 0$ (characteristic of the crossover at $T \approx 10^7$~K).  In each case, the line corresponds to trajectories along which the cooling blob formally reaches $K = 0$ when the blob returns to $r_1$ in a potential with constant circular velocity $v_c$, and cooling time is assumed to be independent of pressure.  Rectangles schematically indicate regimes in which a precipitating system is volatile to feedback fueled by condensation (purple, $v_1/v_c < 1$), marginally susceptible to such feedback (pink, $1 < v_1/v_c < 2$), or quiescent (white, $2 < v_1/v_c$).
\vspace*{1em}
\label{fig-uplift}}
\end{figure*}

Assuming for the time being that the blob's cooling time remains nearly independent of gas pressure during its rise and fall, we can estimate the time of flight necessary for condensation by integrating $d \ln K / dt = - t_{\rm cool}^{-1}$ to obtain
\begin{equation}
  K \approx K_1 \left[ 1 - \frac {6 - 3 \lambda} {5} \frac {\Delta t} {t_{\rm cool}(r_1)} \right]^{5/(6 - 3 \lambda)} \; \; .
\end{equation}
The condition for producing condensation is therefore
\begin{equation}
  \Delta t \; \gtrsim \; \frac {5} {6 - 3 \lambda} \, t_{\rm cool} (r_1) \; \; ,
\end{equation}
or, equivalently, 
\begin{equation}
  \frac {v_1} {v_c} \; \gtrsim \; 	\sqrt{
  				          	 2 \; \ln 
						  \left[ 
				            	  \frac {5} {6 - 3 \lambda} 
   					   	  \frac {t_{\rm ff}(r_{\rm max})} {\Delta t} 
					   	  \left( \frac {t_{\rm cool}} {t_{\rm ff}} \right)_{r_1} 
					  	 \right] 
					          }
					   \; \; .
					   \label{eq-v1vc}
\end{equation}
The right-hand side of this second expression is generally close to unity, as long as $t_{\rm cool} / t_{\rm ff}$ is not orders of magnitude greater than unity.  Thus, the question of condensation depends primarily on the initial velocity of the blob's ballistic trajectory.  Figure~\ref{fig-uplift} shows the critical values of $v_1 / v_c$ required for condensation, according to equation (\ref{eq-v1vc}), as functions of $t_{\rm cool} / t_{\rm ff}$ for $-1 \leq \lambda \leq 0.5$.  Notice that uplift velocities $v_1 \approx v_c$ can induce condensation of blobs that initially have $t_{\rm cool} / t_{\rm ff} \approx 3$ but that $v_1 \gtrsim 1.5 v_c$ is needed for uplift to induce condensation of blobs that initially have $t_{\rm cool} / t_{\rm ff} \gtrsim 10$.  This simple relationship is likely to be the biggest driver of condensation in more complex treatments of multiphase outflows \citep[e.g.,][]{Thompson_2016MNRAS.455.1830T}

\section{Thermal Instability: General Considerations}
\label{sec-GeneralConsiderations}

Uplift is not the only route to condensation. Under certain conditions, a medium in thermal balance can be unstable to condensation when small isobaric entropy perturbations are introduced.  This topic has considerable heritage in the literature, and some of those findings can appear contradictory.  Therefore, in order to clarify the conditions under which condensation can occur, this section goes back to basics to outline what happens to an entropy perturbation in a background medium in which the net cooling rate per unit mass ${\cal L}$ depends on both local thermodynamic conditions and on location within the system.  Here we will use the entropy equation $d \ln K / dt  = - (2 \mu m_p / 3 kT )\, {\cal L} \, $, which is appropriate for a monatomic ideal gas, but the results are similar for more complicated equations of state.

\subsection{Condensation in a Uniform Background}

Without gravity, the pressure $P$ and entropy $K$ of the background medium can be uniform.   The contrast in entropy between the background and a perturbation with entropy $\tilde{K}$ is $\delta \ln K = \ln (\tilde{K}/K)$, and it evolves with time according to 
\begin{equation}
  \frac {d} {dt} ( \delta \ln K ) 
                        \: =  \: - \left( \frac  {2} {3} \frac {\mu m_p} {k} \right) 
                        			\delta \left( \frac {\cal L} {T} \right)
                        \: =  \: - \left( \frac  {2} {3} \frac {\mu m_p} {k} \right)
                                       \left. \frac {\partial ({\cal L}/T)} {\partial \ln K} \right|_A \delta \ln K
                        \; \; ,
\end{equation}
where $A$ is an arbitrary thermodynamic quantity complementary to $K$ that remains constant during the process of condensation.  Spatial gradients of ${\cal L}$ do not appear in this equation because they must vanish in order for the background medium to remain uniform.  The perturbation's amplitude therefore grows monotonically with time if
\begin{equation}
      \left. \frac {\partial ({\cal L}/T) } {\partial \ln K} \right|_A  \: < \: 0 \; \; ,
\end{equation}
which is equivalent to the thermal instability condition originally derived by \citet{Balbus86}.  If the background medium is thermally balanced (${\cal L} = 0$), then this condition reduces to 
\begin{equation}
      \left. \frac { \partial {\cal L} } {\partial \ln K} \right|_A  \: < \: 0 \; \; ,
\end{equation}
which is equivalent to the thermal instability condition originally derived by Field (1965).

To obtain the timescale on which isobaric gas condenses, we set $A = P$, so that 
\begin{equation}
  \frac {d} {dt} ( \delta \ln K )  \: =  \:  \omega_{\rm ti}  ( \delta \ln K )
     \; \; \; \; \; , \; \; \; \; \; 
  \omega_{\rm ti} \equiv - \left( \frac  {2} {3} \frac {\mu m_p} {k} \right)
                                       \left. \frac {\partial ({\cal L}/T)} {\partial \ln K} \right|_P
\end{equation}
Then we separate ${\cal L}$ into a cooling rate per unit mass $C$ and a heating rate per unit mass $H$, so that ${\cal L} = C - H$.  If all of the cooling is radiative, then $C/T \propto K^{-(6-3\lambda)/5} P^{(1+2\lambda)/5}$ and
\begin{equation}
    \omega_{\rm ti} = \frac {6 - 3 \lambda} {5 \, t_{\rm cool}}
   					 \, \left[  \, 1 \, + \, \left( \frac {5} {6 - 3 \lambda} \right) \frac {H} {C} 
                                   		\left. \frac {\partial \ln (H/T)} {\partial \ln K} \right|_P \, \right]
                        \; \; .
\end{equation}
The second term inside the square brackets vanishes if $H=0$.  It also vanishes if heating per unit volume is constant, because in that case $H/T$ is constant in an isobaric medium.  Condensation therefore happens as long as $\lambda < 2$ and the factor in square brackets is positive, and it progresses exponentially on a time scale $\omega_{\rm ti}^{-1}$.  Temporal variations in $H$ have no effect on a perturbation's growth rate if both the perturbation and the ambient medium share the same value of $H/T$ at each moment in time.  However, small entropy perturbations in a medium with no heating (i.e., $H=0$) do not get much of an opportunity to develop into non-linear condensates, because the timescale for perturbation growth is then quite similar to the timescale for cooling of the background medium \citep[e.g.,][]{Malagoli_1987ApJ...319..632M,bs89}.  

\subsection{Condensation in a Gravitationally Stratified Background}

In the presence of gravity, the pressure of a static background medium cannot be uniform, and condensation couples interestingly with buoyancy \citep[e.g.,][]{Defouw_1970ApJ...160..659D,Cowie_1980MNRAS.191..399C,Nulsen_1986MNRAS.221..377N,Malagoli_1987ApJ...319..632M,Loewenstein_1989MNRAS.238...15L,bs89}.  The strength of this coupling depends on the entropy gradient $\nabla \ln K$ in the medium, because
\begin{equation}
  \frac {\partial} {\partial t} ( \delta \ln K ) 
                        \: =  \:  \omega_{\rm ti}  ( \delta \ln K ) +  \omega_{\rm sw}  ( \delta \ln P )
                                    - \dot{\boldxi} \cdot \nabla \ln K 
                          \; \; ,
   \label{eq-Kdot} 
\end{equation}
where ${\boldxi}({\bf r},t) \equiv {\bf r} - {\bf r}_0({\bf r},t)$ represents the displacement of gas at location ${\bf r}$ and time $t$ from its initial location ${\bf r}_0$ at time $t=0$.  In this expression, the quantity   
\begin{equation}
  \omega_{\rm sw} \equiv - \left( \frac  {2} {3} \frac {\mu m_p} {k} \right)
                                       \left. \frac {\partial ({\cal L}/T)} {\partial \ln P} \right|_{K,r}
                             = - \frac {6 + 2 \lambda} {5 t_{\rm cool}} 
     \label{eq-omega_sw}
\end{equation}
is the inverse of the timescale on which sound waves either grow or decay.  The right-hand side of equation (\ref{eq-omega_sw}) applies to a thermally balanced medium in which heating per unit volume is constant at each radius and implies that circumgalactic sound waves tend to be thermally stable, because $\omega_{\rm sw} < 0$ for essentially all environments of interest.  Thermal instability consequently does not develop unless $\omega_{\rm ti} > 0$.  However, subsequent motions resulting from changes in buoyancy can either increase or diminish the perturbation's (Eulerian) entropy contrast $\delta \ln K$, depending on the sign of $\dot{\boldxi} \cdot \nabla \ln K$.

We are primarily interested in systems in which ${\cal L}$ depends explicitly on position, but it is illuminating to consider first the case in which ${\cal L}$ depends on only the local thermodynamic state, as quantified by $K$ and $P$.  In that case, the thermal instability condition $\omega_{\rm ti} > 0$ cannot be satisfied in a hydrostatic medium that is convectively stable and in which sound waves are thermally stable, as emphasized by \citet{bs89}.  In such a medium, $K$ increases with altitude while $P$ decreases with altitude.  Therefore, the condition ${\cal L}(K,P) = 0$ requires $\omega_{\rm ti}$ and $\omega_{\rm sw}$ to have the same sign.  Otherwise, there cannot be a configuration that is simultaneously hydrostatic, convectively stable, and thermally balanced.  Hence, a background configuration with all of three of these desirable stability characteristics {\em does not exist} unless $\omega_{\rm ti} \leq 0$.

An explicit dependence of ${\cal L}$ on position ${\bf r}$ allows hydrostatic, convectively stable configurations to exist for a medium with $\omega_{\rm ti} > 0$ even if sound waves are thermally stable.  In a spherically symmetric environment, a thermally balanced medium in which ${\cal L}$ depends explicitly on position has
\begin{equation}
       \alpha_K \left. \frac {\partial {\cal L}} {\partial \ln K} \right|_{P,r} 
       + \; \alpha_P \left. \frac {\partial {\cal L}} {\partial \ln P} \right|_{K,r}
       + \; \left. \frac {\partial {\cal L}} {\partial \ln r} \right|_{K,P} 
             \: = \: 0
	\; \; ,
	\label{eq-gradK_gradP_gradr}
\end{equation}
where $\alpha_K \equiv {\bf r} \cdot \nabla \ln K$ and $\alpha_P \equiv {\bf r} \cdot \nabla \ln P$.
The physical implications of this relationship are easiest to see if $C$ depends only on $K$ and $P$ while $H/T$ depends only on $r$, giving
\begin{equation}
	\frac {d \ln (H/T)} {d \ln r} 
		\: = \: -  \,  
	 \left( \alpha_K \, \omega_{\rm ti} + {\alpha_P \, \omega_{\rm sw}} \right) \, t_{\rm cool}
	\; \; .
\end{equation}
According to this equation, thermal instability can proceed in a thermally balanced medium satisfying the other three stability conditions as long as the decline in $H/T$ with radius is sufficiently rapid.  

\subsection{Circumgalactoseismology}

From here on, we will focus on media that have $\omega_{\rm ti} > 0$ but in which sound waves are stable ($\omega_{\rm sw} < 0$).  Entropy perturbations can then be thermally unstable and undergo buoyancy-driven motions governed by the momentum equation
\begin{equation}
 	\left( 1 + \delta \ln \rho \right) 
 \ddot{\boldxi} \; = \; 
          \left[ \frac {3} {5} (\delta \ln K) + \frac {2} {5} (\delta \ln P) \right]  \nabla \phi
         		- \frac {P} {\rho} \nabla (\delta \ln P) 
  \; \; .
  \label{eq-ddotxi}
\end{equation}
Here, $\nabla \phi$ represents the gradient of the gravitational potential, and we have assumed $\nabla P = - \rho \nabla \phi$, so that the background state is in hydrostatic equilibrium.  
Differentiating equation (\ref{eq-Kdot}) with respect to time and applying equation (\ref{eq-ddotxi})  after dropping the non-linear $(\delta \ln \rho) \ddot{\boldxi}$ term then gives  
\begin{equation}
 \left[ \, \frac {\partial^2} {\partial t^2} \; 
               - \; \omega_{\rm ti} \frac {\partial} {\partial t} \; + \; \omega_{\rm buoy}^2 \, \right] 
   \left( \delta \ln K \right)
    	\; = \; 
    \left[ \, \omega_{\rm sw} \frac {\partial} {\partial t}  
		\, - \, \frac {2} {3} \omega_{\rm buoy}^2
    		\, + \, \omega_{\rm buoy}^2 \frac {c_s^2} {g} \, \frac {\partial } {\partial r} 
		\, \right] (\delta \ln P) 
    \; \; ,
    \label{eq-Kevol}
\end{equation}
where $\omega_{\rm buoy} \equiv [ (3/5) \nabla \phi \cdot \nabla \ln K]^{1/2} = (6 \alpha_K/5)^{1/2} t_{\rm ff}^{-1}$ is the usual Brunt-V{\"a}is{\"a}l{\"a} frequency for buoyant oscillations and $c_s = (5P/3\rho)^{1/2}$ is the adiabatic sound speed.  Note that this equation has been obtained without using either the continuity equation or the Boussinesq approximation 
and relates internal gravity waves (left-hand side) to the corresponding pressure disturbances (right-hand side). 

If the right-hand side of the equation is negligible, then an entropy perturbation can move through the background medium without dissipative resistance, and the qualitative behavior of the perturbation's (Eulerian) entropy contrast $\delta \ln K$ is obvious.  Entropy perturbations in a medium with $\omega_{\rm ti} \gg \omega_{\rm buoy} > 0$ grow monotonically with time, while those in a medium with $\omega_{\rm buoy} \gg \omega_{\rm ti} > 0$ oscillate with continually increasing amplitude.  The oscillations grow because of a phase difference between $\xi_r \equiv ({\bf r} \cdot {\boldxi})/r$, which is the perturbation's displacement in the radial direction, and the radial restoring force, which is proportional to $\delta \ln K$. During most of each oscillation cycle, the product of $\xi_r$ and $\delta \ln K$ is negative, meaning that buoyancy forces are usually pushing the perturbation back toward the midpoint of the oscillations.  However, $\delta \ln K > 0$ at the beginning of each upward excursion above the midpoint, and $\delta \ln K < 0$ at the beginning of each downward excursion below the midpoint.  The resulting buoyancy impulses cause the amplitudes of both $\boldxi$ and $\delta \ln K$ to increase with time, as originally found by \citet{Defouw_1970ApJ...160..659D}.

Substituting $\delta \ln K \propto \, e^{- i \omega t}$ into equation (\ref{eq-Kevol}) with the right-hand side set to zero gives the frequency solutions 
\begin{equation}
  \omega \; = \; \omega_\pm \; \equiv \; \frac {i \omega_{\rm ti}} {2} 
         \: \pm \: 
          \left( \omega_{\rm buoy}^2 - \frac {\omega_{\rm ti}^2} {4} \right)^{1/2} 
      \; \; .
      \label{eq-omega}
\end{equation}
Low-entropy perturbations in a such a medium progress steadily toward non-linear condensation if $\omega_{\rm buoy} < \omega_{\rm ti} / 2$, which corresponds to 
\begin{equation}
  \alpha_K <  \frac {3(2-\lambda)^2} {40}  \left( \frac {t_{\rm ff}} {t_{\rm cool}} \right)^2
  \; \; ,
\end{equation}
when expressed in terms of $t_{\rm cool}$, $t_{\rm ff}$, and $\alpha_K$.  However, the qualitative behavior of the oscillatory entropy perturbations that arise when $\omega_{\rm buoy} \gg \omega_{\rm ti}$ depends critically on how one treats the $\delta \ln P$ side of the equation.  We will now proceed with a careful look at those $\delta \ln P$ terms, in order to clarify some seemingly contradictory claims in the astrophysical literature on local thermal instability.

\subsubsection{The Bobbing-Blob Approximation}
\label{sec-BobbingBlob}

One intuitively appealing approach is to idealize an entropy perturbation as a discrete, isolated blob in which all of the gas shares the same displacement vector \citep[e.g.,][]{Cowie_1980MNRAS.191..399C,Nulsen_1986MNRAS.221..377N,Loewenstein_1989MNRAS.238...15L}.  In a medium with $\omega_{\rm buoy} \gg \omega_{\rm ti}$, the vertical bobbing of the blob necessarily excites internal gravity waves that propagate away from the blob.  This transfer of kinetic energy from the blob to its environment implies that the blob's motion must experience some resistance in addition to the background pressure  gradient responsible for maintaining the oscillations.  Adding the necessary resistance by replacing the pressure and density perturbation terms in the equation of motion with a heuristic dissipation term, so that 
\begin{equation}
  \ddot{\boldxi} \, = \, \frac {3} {5} (\delta \ln K) \nabla \phi \, - \, \omega_{\rm D} \dot{\boldxi}
  \; \; ,
\end{equation}
qualitatively accounts for those kinetic energy losses by dissipating the blob's kinetic energy on the timescale $\omega_{\rm D}^{-1}$.   This replacement results in the frequency solutions
\begin{equation}
  \omega \; = \; \omega_\pm \; \equiv \; \frac { i (\omega_{\rm ti} - {\omega_{\rm D}})} {2} 
         \: \pm \: 
          \left[ \omega_{\rm buoy}^2 - \frac {(\omega_{\rm ti}+\omega_{\rm D})^2} {4} \right]^{1/2} 
      \; \; ,
\end{equation}
which show that the blob's oscillations decay instead of growing if the timescale for dissipating kinetic energy is shorter than the thermal timescale for pumping the oscillations.

Associating the dissipation term with hydrodynamical drag gives $\omega_{\rm D} \sim | \dot{\boldxi} | / \Delta r$, where $\Delta r$ is the blob's thickness in the vertical direction.  Given this scaling for $\omega_D$, thermal pumping will increase the amplitudes of small oscillations until they saturate at $\delta \ln K \sim \alpha_K ( \omega_{\rm ti} / \omega_{\rm buoy} ) ( \Delta r / r )$.  The largest (Eulerian) density fluctuations, $\delta \ln \rho \sim \alpha_K^{1/2} (t_{\rm ff} / t_{\rm cool})$, are therefore obtained for $\Delta r \sim r$. This analytical estimate of the saturation threshold agrees well with numerical simulations of oscillatory thermal instability in thermally balanced media \citep[e.g.,][]{McCourt+2012MNRAS.419.3319M,Meece_2015ApJ...808...43M}.  However, the route through which it was obtained indicates that the saturation process is inherently non-linear.  Furthermore, the bobbing-blob approximation ignores what happens to the propagating internal gravity waves stimulated by the blob's oscillations.  

\subsubsection{Plane-Wave Perturbations}

Alternatively, one can treat the perturbations as Eulerian plane waves \citep[e.g.,][]{Malagoli_1987ApJ...319..632M,Balbus88,bs89}, but then some care needs to be taken with the physical interpretation.  Making the usual plane-wave assumption that $\delta \ln K$, $\delta \ln P$, and $\boldxi$ are all $\propto \exp [i ({\bf k} \cdot {\bf r} - \omega t) ]$ and defining ${\cal R} \equiv \delta \ln P / \delta \ln K$ allows us to convert equation (\ref{eq-Kevol}) to the dispersion relation
\begin{equation}
  \omega^2 \, - \, i \, (\omega_{\rm ti} + \, {\cal R} \omega_{\rm sw}) \, \omega
     \, - \left( 1 + \frac {2} {3} {\cal R} - i {\cal R} \frac {c_s^2  k_r} {g}  \right) \omega_{\rm buoy}^2
         \, = \,  0
     \; \; .
     \label{eq-disprel}
\end{equation}
In order to account for the influence of pressure perturbations on this dispersion relation through the ratio ${\cal R}$, we need to bring in the continuity equation for the perturbation, which can be expressed as
\begin{equation}
 \frac {5} {3} \left( \nabla \cdot \boldxi \right) \; = \;   
 	\Delta \ln K - \Delta \ln P 
  \label{eq-continuity}
\end{equation}
when written in terms of the Lagrangian perturbation amplitudes $\Delta \ln P = \delta \ln P + \alpha_P (\xi_r / r)$ and $\Delta \ln K \equiv \delta \ln K + \alpha_K (\xi_r / r)$.   Using the plane-wave assumption to simplify the tangential momentum equation, one can write the divergence of the displacement field as
\begin{equation}
  \nabla \cdot \boldxi \, = \, \left( i k_r r  + \nabla \cdot {\bf r} \right) \frac {\xi_r} {r}  
  					- \frac {3 c_s^2 k_\perp^2} {5 \omega^2} \delta \ln P
					\; \; ,
	 \label{eq-divergence}
\end{equation}
where $k_r$ and $k_\perp$ are the radial and tangential wavenumbers, respectively, and the coordinate divergence term is $\nabla \cdot {\bf r} = 2$ in a spherical environment but can be ignored in a plane-parallel approximation.  Combining equations (\ref{eq-continuity}) and (\ref{eq-divergence}) and recognizing that $\alpha_P = - 5 gr / 3 c_s^2$ then leads to
\begin{equation}
  {\cal R} \, = \, \frac {\omega^2} {\omega^2 - c_s^2 k_\perp^2}
		\left[ \frac {5} {3 \alpha_K} 
				\left( i k_r r + \nabla \cdot {\bf r} - \frac {gr} {c_s^2} \right)
			\left( 1 - \frac {\Delta \ln K} {\delta \ln K} \right)
		      + \frac {\Delta \ln K} {\delta \ln K} \right]
	\; \; .
  \label{eq-R}
\end{equation}
Applying the plane-wave assumption to equation (\ref{eq-Kdot}) gives the ratio of Lagrangian to Eulerian entropy perturbation amplitudes:
\begin{equation}
 \frac  {\Delta \ln K} {\delta \ln K} 
 	\, = \, i \left( \frac {\omega_{\rm ti} + {\cal R} \omega_{\rm sw}} {\omega} \right)
  \; \; .
   \label{eq-DeltaK_deltaK}
\end{equation}
The dispersion relation for perturbations with short radial wavelengths ($ | k_r r |  \gg 1$) therefore reduces to
\begin{equation}
  \omega^2 \, - \, i \, (\omega_{\rm ti} + \, {\cal R} \omega_{\rm sw}) \, \omega
     \, - \left( 1 + \frac {2} {3} {\cal R} \right) 
     	  \left( \frac {c_s^2 k_\perp^2 - \omega^2} {c_s^2 k^2 - \omega^2} \right)
		\omega_{\rm buoy}^2 
         \, = \,  0
     \; \; .
     \label{eq-short_disprel}
\end{equation}
where $k^2 \equiv k_r^2 + k_\perp^2$.

The literature on thermal instability in circumgalactic media has generally focused on short-wavelength modes with $\omega^2 \ll c_s^2 k_\perp^2 /  | k_r r | $, which are nearly pure internal gravity waves.  In that case, the amplitude ratio ${\cal R} = \delta \ln P / \delta \ln K$ is small, and the dispersion relation reduces further to 
\begin{equation}
  \omega^2 \, - \, i \, \omega_{\rm ti} \omega
     \, - \frac {k_\perp^2} {k^2} \omega_{\rm buoy}^2 
         \, = \,  0
     \label{eq-slow_disprel}
\end{equation}
\citep[e.g.,][]{Malagoli_1987ApJ...319..632M,bnf09}.  For these modes, Eulerian pressure perturbations influence the gravity-wave oscillator only through the pressure-gradient term in equation (\ref{eq-Kevol}).  The resulting $k_\perp^2 / k^2$ factor slows the oscillations because the gravitational force correcting a given radial displacement meets with greater inertial resistance from transverse displacements as $k_\perp^2 / k^2$ declines \citep[e.g.,][]{bnf09}.  As a consequence, the critical value of $\alpha_K$ separating monotonic perturbation growth from oscillatory growth for a given $t_{\rm cool} / t_{\rm ff}$ ratio becomes larger than one would obtain by setting $\delta \ln P = 0$.  

Some treatments in the literature \citep[e.g.,][]{bs89,McCourt+2012MNRAS.419.3319M} arrive at a similar dispersion relation by setting $\nabla \cdot \boldxi = 0$.  With that restriction, we find
\begin{equation}
  {\cal R} = \frac {\alpha_P} {\alpha_K}  \left( 1 - \frac {\Delta \ln K} {\delta \ln K} \right) 
  		+ \frac {\Delta \ln K} {\delta \ln K}
	\; \; ,
\end{equation}
and obtain the dispersion relation
\begin{equation}
  \omega^2 \, - \, i \, (\omega_{\rm ti} + \, {\cal R} \omega_{\rm sw}) \, \omega
     \, - \left( 1 + \frac {2} {3} {\cal R} \right) 
     	  \left[ \frac {k_\perp^2 r^2} {k^2 r^2 + i k_r r (\nabla \cdot {\bf r} ) } \right]
		\omega_{\rm buoy}^2 
         \, = \,  0
     \label{eq-no_div_disprel}
\end{equation}
without having to assume $ | k_r r |  \gg 1$.   If the $i k_r r (\nabla \cdot {\bf r})$ term can be neglected, then this relation would reduce to equation (\ref{eq-slow_disprel}) if ${\cal R}$ were small. However, $|{\cal R}|$ is typically of order unity for divergence-free modes, indicating that they are qualitatively different from the internal gravity waves discussed in the previous paragraph.  In fact, the restriction $\nabla \cdot \boldxi = 0$ implies that the radial wavenumber for such modes is imaginary, with
\begin{equation}
  k_r r = i \left[ k_\perp^2 r^2  
  		\left( \frac {c_s^2} {gr} \right)
  		\frac {\omega_{\rm buoy}^2} {\omega^2} 
		+ ( \nabla \cdot {\bf r} ) \right]
\end{equation} 
in the adiabatic limit.  These divergence-free modes are therefore evanescent in the radial direction.  Furthermore, the associated growth rate of thermal instability is $\omega_{\rm ti} + (\alpha_P / \alpha_K) \omega_{\rm sw}$, which exceeds $\omega_{\rm ti}$ in media with stable sound waves (i.e., $\omega_{\rm sw} < 0$), because $\alpha_P/\alpha_K < 0$.

\subsubsection{Onset of Non-linearity}
\label{sec-nonlinearity}

Prior analyses of circumgalactic thermal instability did not focus on the consequences of modes in which $\omega_{\rm sw}$ is important, but here we would like to understand why the growth of oscillatory thermal instability saturates with $\delta \ln \rho \sim \alpha_K^{1/2} (t_{\rm ff}/t_{\rm cool})$, as predicted by the heuristic bobbing-blob model and as observed in numerical simulations \citep{McCourt+2012MNRAS.419.3319M,Meece_2015ApJ...808...43M}.  If only low-${\cal R}$ or divergence-free modes are permitted, then there is no channel (other than viscosity) for dissipation of the kinetic energy that oscillatory thermal instability introduces into the system.  Non-linear mode coupling can exchange energy among those gravity-wave modes but does not by itself limit the overall growth of a set of modes that are all thermally unstable.  In order to achieve a steady state in which $\delta \ln \rho$ depends on $t_{\rm cool}$, non-linear mode coupling must instead channel the accumulating gravity-wave energy into acoustic modes, which damp through radiative losses on a timescale $\sim | \omega_{\rm sw}^{-1} | \sim t_{\rm cool}$.  

In lieu of presenting a complete solution to that mode coupling problem in this paper we would instead like to briefly illustrate how the gravity-wave saturation scale $\delta \ln \rho \sim \alpha_K^{1/2} (t_{\rm ff}/t_{\rm cool})$ arises from non-linearity.  We are interested in knowing how mode coupling changes the amplitudes $A_i$ of the radial-displacement eigenmodes $\xi_i = A_i \exp [ i({\bf k}_i \cdot {\bf r} - \omega_i t)]$, for which $\omega_i ({\bf k}_i)$ satisfies equations (\ref{eq-disprel}), (\ref{eq-R}), and (\ref{eq-DeltaK_deltaK}). The full radial momentum equation for a superposition of those modes can be expressed as
\begin{equation}
 (1 + \delta \ln \rho) \, \left( \sum_i \ddot{\xi}_i \right) \; = \; - \sum_j \omega_j^2 \xi_j
  \; \; ,
  \label{eq-full_ddotxi}
\end{equation}
and the amplitudes $A_i$ remain constant if the nonlinear terms are ignored.  After canceling the linear terms, we can rewrite this equation in a form that relates the rates of amplitude change to those non-linear terms,
\begin{equation}
  \sum_i \left[ \left( \frac{ \ddot{A}_i - i 2 \omega_i \dot{A}_i } {A_i} \right) \xi_i \right] 
         \; = \; 
  	- \left( \sum_j \omega_j^2 \xi_j \right)
	  \left[ \sum_{n = 1}^{\infty} \left( - \, \delta \ln \rho \right)^n \right] 
        \; \approx \; 
  	- \left( \sum_j \omega_j^2 \xi_j \right)
	  \left[ \sum_m \left( \delta \ln \rho \right)_m \right] 
  \; \; ,
  \label{eq-triads}
\end{equation}
where $(\delta \ln \rho)_m$ represents the oscillating density perturbation corresponding to eigenmode $\xi_m$, and we have kept only the lowest-order non-linear terms on the right-hand side.  Resonant energy transfer can therefore happen among triads of modes that satisfy ${\bf k}_i = {\bf k}_j + {\bf k}_m$ and $\omega_i = \omega_j + \omega_m$.  

Here, we are most interested in coupling between a typical internal gravity wave of wavenumber $k$ and frequency $\omega_g \approx \, (k_\perp/k) \omega_{\rm buoy}$ and a pair of sound waves with frequencies $\sim c_s k$ that differ by $\omega_g$ so that their beat pattern resonates with the gravity wave.  In a purely adiabatic system, this triad of waves would eventually come into energy equipartition with $A_g \omega_g   \sim A_s c_s k$, where $A_g$ and $A_s$ are the gravity-wave and sound-wave displacement amplitudes, respectively.  On the right-hand side of equation (\ref{eq-triads}), terms corresponding to pairs of sound waves of appropriate wavelength have magnitudes $\sim (A_s c_s k)^2 k$ because $( \delta \ln \rho )_m \sim A_s k_m$.  In a system in equipartition, these terms are $\sim (A_g \omega_g)^2 k$, implying that mode coupling transfers energy from sound waves into gravity waves and back again at a rate $\sim A_g \omega_g k$.

The system we are considering is not adiabatic.  Instead, gravity waves grow in amplitude on a timescale $\sim \omega_{\rm ti}^{-1}$ until their amplitudes are large enough for mode coupling to channel their energy into dissipative sound waves on the same timescale.  This happens when 
\begin{equation}
   \frac {A_g} {r} \sim \frac {\omega_{\rm ti}} {\omega_{\rm buoy}} \frac {1} {k r} 
   	\; \; \; \; \; , \; \; \; \; \;
    ( \delta \ln \rho )_g
      \; \sim \; \delta \ln K  
      \; \sim \; \alpha_K \frac {A_g} {r}  
      \; \sim \; \alpha_K \frac {\omega_{\rm ti}} {\omega_{\rm buoy}} \frac {1} {k r}
      \; \sim \; \alpha_K^{1/2} \frac {t_{\rm ff}} {t_{\rm cool}} \frac {1} {k r}
    \; \; ,
\end{equation}
where $( \delta \ln \rho )_g$ is the density perturbation scale associated with gravity-wave oscillations.  

\subsection{Buoyancy Damping}
\label{sec-BuoyancyDamping}

The saturation amplitude derived from non-linear mode coupling considerations in \S \ref{sec-nonlinearity} closely resembles the one derived from the heuristic bobbing-blob model in \S \ref{sec-BobbingBlob}.  In the bobbing-blob case, the timescale for energy transfer from the blob to the surrounding medium is $\sim (A_g \omega_{\rm buoy})^{-1} \Delta r$, and energy is permanently lost if it propagates away from the blob without contributing to the restoring force on it.  Analogously, the timescale for energy transfer from internal gravity waves to damped acoustic modes is $\sim (A_g \omega_{\rm buoy})^{-1} k^{-1}$, regardless of the mechanism that dissipates sound-wave energy.  In both cases, the saturation amplitudes are largest for inhomogeneities with $kr \sim 1$.

Saturation of thermally unstable gravity-wave oscillations therefore has important consequences for condensation in the circumgalactic medium.  In an otherwise static medium with $\alpha_K \gg (t_{\rm ff} / t_{\rm cool})^2$, energy transfer from buoyant oscillations into dissipative modes prevents a gas blob with slightly lower entropy than its surroundings from condensing.  At first, the blob descends faster than it can radiate away its thermal energy and accelerates until it passes through a layer of equivalent entropy.  After it overshoots that layer it begins to bob, and its subsequent behavior depends on the amplitude $|\delta \ln K|$ of its Eulerian entropy contrast.  If that amplitude is less than the saturation amplitude, then thermal pumping will cause it to grow to the saturation amplitude.  But if $|\delta \ln K |$ is greater than the saturation amplitude, then coupling of buoyant oscillations to acoustic modes causes $|\delta \ln K |$ to decay until it reaches the saturation amplitude.  Hereafter, we will refer to such a decline in perturbation amplitude as {\em buoyancy damping}, because it results from damping of buoyant oscillations in a gravitationally stratified medium. 

In contrast, buoyancy does not inhibit condensation in a medium with $\alpha_K \ll (t_{\rm ff} / t_{\rm cool})^2$, because there are no oscillations. Instead, low-entropy blobs proceed steadily toward becoming non-linear condensates.  In fact, the onset of non-linearity can actually {\em assist} perturbation growth, because it inhibits the descent of low-entropy gas toward layers of equivalent entropy (e.g., Nulsen 1986).  Therefore, the condition
\begin{equation}
  \alpha_K \sim \left( \frac {t_{\rm ff}} {t_{\rm cool}} \right)^2 
\end{equation}
corresponds to a critical entropy-profile slope separating media in which linear thermal instability leads to condensation from media in which linear thermal instability saturates at relatively small amplitudes.\footnote{\citet{bnf09} arrive at a similar-looking criterion, their equation (10), via different reasoning.  Instead of arguing that oscillatory perturbations saturate at a relatively small amplitude when $t_{\rm cool} \gg t_{\rm ff}$, they instead state without proof that radiative cooling will not have a big impact on gravity-wave oscillations that spend half of their time overdense and cool and the other half of it underdense and warm.}  

At the margin between these regimes, where $\omega_{\rm buoy} \sim \omega_{\rm ti}$, thermal instability itself starts to reduce the frequency of gravity-wave oscillations, which slows the rate of energy transfer into sound waves.  In a system with $\alpha_K \gtrsim (t_{\rm ff} / t_{\rm cool})^2$, the saturation scale for thermally unstable gravity-wave oscillations with $kr \sim 1$ becomes
\begin{equation}
   \frac {A_g} {r} \sim \frac {\omega_{\rm ti}} {\omega_{\rm buoy}} 
   			                \left( 1 - \frac {\omega_{\rm ti}^2} {4 \omega_{\rm buoy}^2} \right)^{-1/2} 
   	\; \; \; \; \; , \; \; \; \; \;
    ( \delta \ln \rho )_g
      \; \sim \; \alpha_K \frac {A_g} {r}  
      \; \gtrsim \; \frac {t_{\rm ff}^2} {t_{\rm cool}^2} 
                        \left( 1 - \frac {\omega_{\rm ti}^2} {4 \omega_{\rm buoy}^2} \right)^{-1/2} 
    \; \; .
\end{equation}
In other words, there is a lower limit of $(\delta \ln \rho)_g \sim (t_{\rm ff} / t_{\rm cool})^2$ on the saturation amplitude of gravity-wave perturbations, and the limiting amplitude approaches unity as $\omega_{\rm buoy}$ approaches $\omega_{\rm ti} / 2$ from above.  A reduction of the entropy gradient to $\alpha_K \sim (t_{\rm ff} / t_{\rm cool})^2$ consequently eliminates buoyancy damping and allows thermal instability to proceed to condensation.\footnote{After submitting this paper, we recognized that turbulence can alter the condensation criterion by interfering with buoyancy damping.  Turbulence enters the picture when the turbulent velocity dispersion is similar in magnitude to the buoyancy-driven motions that result in damping of thermal instability, and it can enable condensation by levitating low-entropy blobs that would otherwise descend in a medium with $\alpha_K > (t_{\rm ff} / t_{\rm cool})^2$.  We will explore more deeply how turbulence alters the condensation criterion in a separate paper. For a complementary viewpoint on the role of turbulence in promoting condensation, see \citet{Gaspari_2017MNRAS.466..677G}. }

\newpage

\section{Thermal Instability in Idealized Numerical Simulations} 
\label{sec-TI_NumericalSimulations}

The analysis in \S \ref{sec-GeneralConsiderations} provides a conceptual foundation for interpreting the conditions governing thermal instability and condensation in numerical simulations of circumgalactic media.  \citet{McCourt+2012MNRAS.419.3319M} showed that thermal instability in the oscillatory regime does indeed saturate at relatively small perturbation amplitudes, while the simulations of \citet{Meece_2015ApJ...808...43M} show something more:  Condensation and feedback in one part of a globally balanced system can alter the behavior of thermal instability elsewhere in the system by changing the large-scale slope of its entropy profile.  This section applies the results of \S \ref{sec-GeneralConsiderations} to show how that happens.

Before proceeding, we want to emphasize that this connection between thermal instability and the large-scale entropy gradient turns out to be critical for understanding how feedback regulates condensation, star formation, and black-hole accretion.  Most critically, it implies that central injection of thermal energy inevitably promotes condensation because it reduces the central entropy gradient.  Therefore, if feedback energy is to suppress condensation, it most effectively does so when distributed in ways that {\em increase} the system's entropy gradient, as we will show in \S \S \ref{sec-GlobalBalance},\ref{sec-Illustrations}.

\subsection{Buoyancy Damping in Simulations}
\label{sec-buoyancy_damping_sims}

\citet{McCourt+2012MNRAS.419.3319M} drew attention to the dependence of saturation amplitude on $t_{\rm cool} / t_{\rm ff}$ using simulations of a plane-parallel medium in which heating balances cooling within each equipotential layer.  They showed that small entropy perturbations introduced into an initially isothermal medium grow to an Eulerian entropy contrast $\sim \alpha_K (\omega_{\rm ti} / \omega_{\rm buoy})$ before saturating.  The interpretation they offered for this saturation was that non-linear mode coupling leads to dissipation when $\xi_r / r \sim \omega_{\rm ti} / \omega_{\rm buoy}$, but they were not specific about the nature of the dissipation or the mechanism through which mode coupling disrupts the growth of gravity-wave oscillations.  They concluded that the criterion $t_{\rm cool} / t_{\rm ff} \lesssim 1$ was a necessary condition for condensation in a plane-parallel medium.

\begin{figure*}[t]
\begin{center}
\includegraphics[width=6.7in, trim = 0.0in 0.0in 0.0in 0.0in]{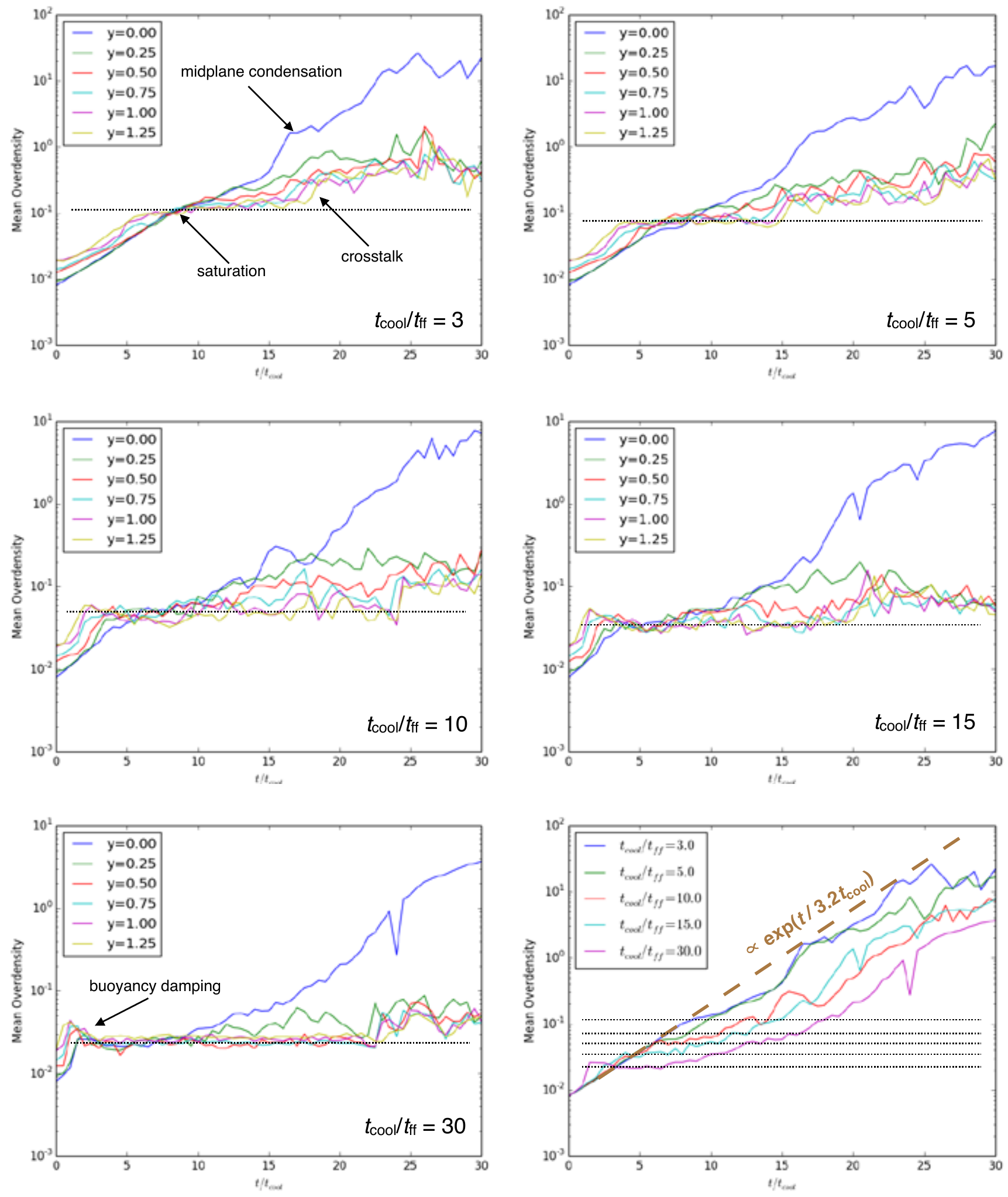} \\
\end{center}
\caption{ \footnotesize 
Growth of the rms amplitudes of density perturbations through thermal instability in thermally balanced plane-parallel simulations.  Each panel shows perturbation growth as a function of $t/t_{\rm cool}$ in the simulations of \citet{Meece_2015ApJ...808...43M}, in which average heating equals average cooling in each equipotential layer.  The simulations all begin with nearly isentropic initial conditions within a pressure scale height of the midplane, and labels in five of the panels show the initial value of $t_{\rm cool} / t_{\rm ff}$ in the simulation domain.  Lines of different colors in those panels show $\langle \Delta \rho / \rho \rangle_{\rm rms}$ at different heights above the midplane, where $y$ is the height in units of the initial pressure scale height; the blue line shows condensation at the midplane ($y=0$).  
Black dotted lines in those five panels show the level at which perturbations saturate because of buoyancy damping.   The sixth panel (lower right) shows the time required for each simulation to progress toward midplane condensation (i.e., Mean Overdensity $\approx 1$).  The five black dotted lines in that panel are the same as those in the other five panels, and a brown dashed line extrapolates the exponential growth rate that precedes saturation.  When midplane condensation happens, the compensating release of heat produces buoyant bubbles (convective crosstalk) that amplify the perturbation amplitudes in higher-lying layers.
\vspace*{1em}
\label{fig-Meece_condensation}}
\end{figure*}

\citet{Meece_2015ApJ...808...43M} performed similar simulations showing that the condensation criterion is more complex, in that condensation is always possible near the midplane, even in systems with $t_{\rm cool} / t_{\rm ff} \gg 1$, if enough time is allowed \citep[see also][]{ChoudhurySharma_2016MNRAS.457.2554C}.  Figure~\ref{fig-Meece_condensation} shows some previously unpublished results from the Meece et al. effort.  Perturbation growth far from the midplane saturates as in \citet{McCourt+2012MNRAS.419.3319M} but is not fully suppressed near the midplane.  Instead, it is significantly delayed for large values of $t_{\rm cool} / t_{\rm ff} \gg 1$ and eventually generates crosstalk that affects perturbation growth in higher-lying layers, for reasons we will now proceed to analyze.

\subsection{Midplane Condensation}
\label{sec-Midplane}

Condensation is always possible near the midplane of a macrophysically balanced system in a plane-parallel potential well because buoyancy damping is ineffective there as long as the gas density does not become infinite at the midplane.  In the bobbing-blob approximation of \S \ref{sec-BobbingBlob}, the boundary between monotonic growth and oscillatory thermal instability is at $\omega_D = \omega_{\rm ti}$, implying that the condition for condensation is
\begin{equation}
  \alpha_K < \frac {3(2-\lambda)^2} {10}  \left( \frac {t_{\rm ff}} {t_{\rm cool}} \right)^2  \; \; .
\end{equation}
In order to determine the zone around the midplane in which condensation can proceed, one can adopt a linear approximation to the potential, in which the freefall time $t_{\rm ff,0}$ is constant with height.  The pressure and entropy at the midplane are $P_0$ and $K_0$, respectively, giving a midplane density $\rho_0 = \mu m_p (n/n_e)^{2/5} (P_0/K_0)^{3/5}$.   To lowest order in height $z$ above the midplane, the pressure profile is
$P(z) = P_0 [ 1 - (z/z_P)^2 ]$, where $z_P \equiv t_{\rm ff,0}(P_0 / \rho_0)^{1/2}$, and the linearized entropy profile can be expressed as $K(z) = K_0 [ 1 + (z/z_K)]$.  The condensation condition in this zone is therefore
\begin{equation}
  \frac {z} {z_K} \; \lesssim \; \frac {3(2-\lambda)^2} {10} 
  				\left[ \frac {t_{\rm ff,0}} {t_{\rm cool,0}}  \right]^2
  				\left[ 1 + \left( \frac {z} {z_K} \right) \right]^{6(2- \lambda)/5}
  				\left[ 1 - \left( \frac {z} {z_P} \right)^2 \right]^{2(1+2 \lambda)/5}		
			    \; \; ,
\end{equation}
where $t_{\rm cool,0} \equiv t_{\rm cool}(K_0,P_0)$. In other words, the thickness of the midplane region that is unstable to condensation is approximately $(t_{\rm ff} / t_{\rm cool})^2$ times the entropy scale height in media with $t_{\rm cool} / t_{\rm ff} \gg 1$.  

\subsection{Isentropic Initial Conditions}

At first glance, the susceptibility to condensation of a system in which $\alpha_K \lesssim (t_{\rm ff} / t_{\rm cool})^2$ would appear to be in conflict with the plane-parallel simulations of \citet{McCourt+2012MNRAS.419.3319M} that start with isentropic initial conditions ($\alpha_K = 0$) and fail to produce non-linear density contrasts.  In those simulations, the fractional fluctuations in entropy and density at one pressure scale height above the midplane saturate at $\sim (t_{\rm ff} / t_{\rm cool})^2$ and remain at this amplitude for many cooling times.  The reason for saturation of perturbation growth in what would seem to be an isentropic medium is that buoyant migration of the initial perturbations causes an entropy gradient to develop.  As lower-entropy fluctuations sink and higher-entropy fluctuations rise, the resulting radial sorting of entropy perturbations increases $\alpha_K$.  Eventually the entropy gradient exceeds $\alpha_K \sim (t_{\rm ff} / t_{\rm cool})^2$, and the condensation condition is no longer satisfied.  The system then transitions to the bobbing-blobs regime, in which (Eulerian) entropy fluctuations saturate at a fractional amplitude $\sim \alpha_K (\omega_{\rm ti} / \omega_{\rm buoy} )$, which corresponds to $\sim (t_{\rm ff} / t_{\rm cool})^2$ for $\alpha_K \sim (t_{\rm ff} / t_{\rm cool})^2$, as shown in \S \ref{sec-BuoyancyDamping}.

\citet{McCourt+2012MNRAS.419.3319M} recognized that development of an entropy gradient through convective sorting of the initial perturbations was responsible for suppressing perturbation growth away from the midplane but did not fully explore the consequences for stability close to the midplane.  Instead, they suppressed both heating and cooling near the midplane.  In contrast, \citet{Meece_2015ApJ...808...43M} chose to allow heating and cooling to proceed near the midplane, and those simulations ended up displaying qualitative behavior distinctly different from the initially isentropic \citet{McCourt+2012MNRAS.419.3319M} simulations, even though they started with nearly isentropic initial conditions within a pressure scale height of the midplane.

Midplane condensation always occurs in the \citet{Meece_2015ApJ...808...43M} simulations, no matter what the initial value of $t_{\rm cool} / t_{\rm ff}$, because there is always a zone around the midplane in which $\alpha_K \lesssim (t_{\rm cool} / t_{\rm ff})^2$.  However, buoyancy damping delays condensation by a time interval depending on $t_{\rm cool} / t_{\rm ff}$.  Without gravity and buoyancy damping, the initial entropy perturbations of \citet{Meece_2015ApJ...808...43M} would experience monotonic exponential growth.  Nearly exponential growth is indeed seen in simulations with $t_{\rm cool} / t_{\rm ff} \lesssim 3$, but larger values of $t_{\rm cool} / t_{\rm ff}$ lead to saturation of perturbation growth at progressively smaller amplitudes (see Figure~\ref{fig-Meece_condensation}).  Buoyancy damping suppresses perturbation growth away from the midplane but succeeds only in delaying it near the midplane.  Delays occur because buoyancy damping limits the initial amplitudes of the seed perturbations that enter the isentropic zone and ultimately condense.  As a result, the time required for condensation at the midplane is a larger multiple of $t_{\rm cool}$ in systems with larger values of $t_{\rm cool} / t_{\rm ff}$.

\subsection{Convective Crosstalk}

Once condensation begins at the midplane in the \citet{Meece_2015ApJ...808...43M} simulations, the idealized algorithm for enforcing thermal balance within each layer ends up stimulating perturbation growth at greater altitudes.  As condensates form near the midplane, the requirement for heating to balance cooling in each layer causes compensating hot bubbles to form.  Those hot bubbles then rise out of the condensation zone.  When they reach higher layers, the bubbles catalyze additional condensation, because regions that were formerly at the mean entropy for their layer, and therefore close to thermal balance, suddenly find themselves significantly below the new mean entropy level.  The net cooling rate in lower-entropy regions of the layer must therefore increase in order to satisfy the thermal balance constraint.  As a result, convective crosstalk between layers causes non-linear perturbation growth to propagate beyond the thermally unstable midplane zone. The same effect was also found in simulations by \citet{ChoudhurySharma_2016MNRAS.457.2554C}.

Does this crosstalk phenomenon, which arises in \citet{Meece_2015ApJ...808...43M} and \citet{ChoudhurySharma_2016MNRAS.457.2554C} from a highly idealized representation of global thermal balance, have a qualitative counterpart in more realistic treatments of feedback?  We suggest that the answer to this question may be yes, because the outflow of feedback energy from a central source in a real galactic system is likely to propagate primarily through the lowest-density, highest-entropy ambient gas as it follows the path of least resistance to greater altitudes.  Unless some mechanism (such as thermal conduction) exists for preferentially channeling feedback energy into denser, cooler gas clumps, then lower-entropy regions will inevitably cool and condense while feedback energy flows outward around them, even if heating and cooling are in global balance.

This phenomenon illustrates how centrally injected thermal feedback renders the surrounding region vulnerable to condensation:  Convection eliminates the central entropy gradient and therefore shuts off buoyancy damping of perturbation growth.  Consequently, centrally injected thermal feedback actually {\em promotes} precipitation rather than suppressing it.  This is why centrally injected thermal feedback fails so miserably to prevent condensation in the cluster-core simulation of \citet{Meece_2016arXiv160303674M} that relies on pure thermal feedback, as well as many other such simulations that rely on central thermal feedback alone.  
With that thought in mind, we now turn our attention from local thermal instability to its implications for global thermal balance.

\section{Evolution of Systems in Global Thermal Balance}
\label{sec-GlobalBalance}

The analyses of \S \ref{sec-GeneralConsiderations} and \S \ref{sec-TI_NumericalSimulations} show that $t_{\rm cool}/t_{\rm ff} \approx 10$ is not a general threshold for condensation.  So why is the lower limit $t_{\rm cool}/t_{\rm ff} \gtrsim 10$ so prevalent among multiphase systems in both numerical simulations \citep[e.g.,][]{Sharma_2012MNRAS.420.3174S,Gaspari+2012ApJ...746...94G,Gaspari+2013MNRAS.432.3401G,Li_2015ApJ...811...73L} and in nature \citep[e.g.,][]{McCourt+2012MNRAS.419.3319M,VoitDonahue2015ApJ...799L...1V,Voit_2015Natur.519..203V,Voit+2015ApJ...803L..21V}?  This section examines how $\min (t_{\rm cool}/t_{\rm ff})$ evolves in systems constrained to be in global thermal balance. It shows that if each individual layer is restricted to remain in thermal balance, then buoyancy damping substantially slows the process of condensation in simulated galactic systems as $\min (t_{\rm cool}/t_{\rm ff})$ rises, causing it to stall near $\min (t_{\rm cool}/t_{\rm ff}) \approx 10$ in simulated cluster cores.  However, if the thermal balance restriction is relaxed in individual layers, then a globally balanced system evolves toward a value of $\min (t_{\rm cool}/t_{\rm ff})$ that depends on the spatial distribution of heat input.  In that case, global stability near $\min (t_{\rm cool}/t_{\rm ff}) \approx 10$ can be realized if heat input exceeds cooling in the inner and outer parts of the system but falls short of matching cooling at intermediate altitudes.

\subsection{Unmixing at Constant Pressure}
\label{sec-unmixing}

From a thermodynamic perspective, development of a multiphase medium under conditions of global thermal balance is essentially the opposite of thermal mixing in a system with constant total energy.  For example, consider a system in which a small fraction of the gas condenses by radiating an amount of heat energy $\delta Q$.  To maintain strict thermal balance, the rest of the gas in the system must gain the same amount of heat energy through feedback. If this condensation and feedback process proceeds at constant pressure, then the volume of the overall system does not change, but its overall entropy must {\em decrease}, because hotter gas gains heat energy while colder gas radiates it away.  This inhomogeneous condensation process is prohibited in a closed system by the second law of thermodynamics.  However, the supply of free energy that is added to balance cooling in a globally balanced system enables it to ``unmix" via condensation.  In this context, feedback is not only necessary to keep the system in overall thermal balance, it is also {\em required} to transform the original single-phase medium into a multiphase medium via thermal instability.

\subsection{Instability in Thermally Balanced Equipotential Layers}

Gravity can suppress ``unmixing'' by causing buoyancy damping.  The simulation results discussed in \S \ref{sec-TI_NumericalSimulations} show that buoyancy damping in a gravitationally stratified medium causes thermal instability to saturate in layers with $\alpha_K \gg (t_{\rm ff}/t_{\rm cool})^2$ and can substantially delay condensation even if $\alpha_K \ll (t_{\rm ff}/t_{\rm cool})^2$.  Extrapolating these findings to a cluster-core environment suggests that a thermally-balanced cluster core with $t_{\rm cool} \lesssim 1$~Gyr and $\min ( t_{\rm cool} / t_{\rm ff}) \sim 10$ may be able to resist condensation for $\gtrsim 5$~Gyr.

The idealized cluster-core simulations of \citet{Sharma_2012MNRAS.420.3174S}, which strictly maintain thermal balance in each equipotential layer, appear to corroborate this suggestion.  In their fiducial simulation (run C10), the core configuration begins with central entropy $\approx 10 \, {\rm keV \, cm^2}$ (corresponding to $t_{\rm cool} \approx 200$~Myr for the adopted cooling function) and $\min ( t_{\rm cool} / t_{\rm ff}) \approx 10$.  Even though the central cooling time is $\ll 1$~Gyr, the system remains almost static for $\approx 2$~Gyr.  Then a large condensation event raises the central entropy to $\sim 20 \, {\rm keV \, cm^2}$, which corresponds to $\min (t_{\rm cool}) \approx 600 \, {\rm Myr}$ and $\min ( t_{\rm cool} / t_{\rm ff}) \approx 15$.  Afterward, the system remains relatively stable for $\sim 3$~Gyr.  The initial delay time before condensation ($\sim 10 \, t_{\rm cool}$) is consistent with a period of buoyancy damping followed by steady exponential growth, as illustrated in Figure~\ref{fig-Meece_condensation}.  

A similar interpretation suits run C1 from \citet{Sharma_2012MNRAS.420.3174S}, which starts with central entropy $\approx 1 \, {\rm keV \, cm^2}$ and $\min ( t_{\rm cool} / t_{\rm ff}) \approx 1$.  Condensation begins promptly in that simulation and raises the ambient specific entropy at small radii to $\sim 10 \, {\rm keV \, cm^2}$ in $\lesssim 2$~Gyr, at which time $\min(t_{\rm cool}) \sim 200 \, {\rm Myr}$ and $\min( t_{\rm cool} / t_{\rm ff}) \sim 10$.  Condensation then subsides, and the entropy and cooling-time profiles remain relatively stable for the next $\sim 2$~Gyr, presumably for the same reason that run C10 resisted condensation for the first 2~Gyr.

In contrast, the transition to $\min ( t_{\rm cool} / t_{\rm ff}) \gtrsim 10$ requires only $\sim 40$~Myr for the \citet{Gaspari+2013MNRAS.432.3401G} simulations, which also enforce thermal balance within equipotential layers.   There are three reasons for the action to proceed on a shorter timescale than in \citet{Sharma_2012MNRAS.420.3174S}. First, the greater metallicity (i.e. solar values) and lower temperature compared with \citet{Sharma_2012MNRAS.420.3174S} result in a shorter central cooling time at a given central entropy level.  Second, the gravitational potential is deeper because of the gravity of the central galaxy's stars and a central supermassive black hole are both included, meaning that the freefall time at each radius is shorter.  Third, and most importantly, \citet{Gaspari+2013MNRAS.432.3401G} continually drive turbulence, which counteracts the effects of buoyancy damping and raises the amplitude of the seed perturbations that thermal instability amplifies.  

Taken together, the simulations of \citet{Sharma_2012MNRAS.420.3174S} and \citet{Gaspari+2013MNRAS.432.3401G} have been viewed as support for the suggestion that systems in global thermal balance naturally stabilize near $\min(t_{\rm cool}/t_{\rm ff}) \approx 10$, as observed in more realistic simulations of cluster cores regulated by precipitation-fed bipolar outflows \citep{Gaspari+2012ApJ...746...94G,Li_2015ApJ...811...73L,Prasad_2015ApJ...811..108P, Meece_2016arXiv160303674M,YangReynolds_2016arXiv160501725Y}.  However, both the analysis of \S \ref{sec-TI_NumericalSimulations} and the analytical toy model for configuration changes presented in the Appendix show that there is nothing unique about $t_{\rm cool}/t_{\rm ff} \approx 10$ from the point of view of local thermal stability.  Instead, the \citet{Sharma_2012MNRAS.420.3174S} simulations remain near $\min(t_{\rm cool}/t_{\rm ff}) \approx 10$ for long time periods because increases in $t_{\rm cool}/t_{\rm ff}$ stretch the time interval between condensation events to multiple Gyr, even though $\min(t_{\rm cool})$ remains $\lesssim 1$~Gyr.

\subsection{Global Thermal Balance with Distributed Heating}
\label{sec-GlobalThermalBalance}

Relaxing the requirement that each equipotential layer remain in thermal balance while retaining the global thermal balance constraint leads to a somewhat different interpretation for the minimum value of $t_{\rm cool}/t_{\rm ff}$.  When this restriction is relaxed, the entropy profile of a globally balanced system can change with time in response to the spatial distribution of heat input.  In systems that can be divided into an inner isentropic zone and an outer power-law zone (see Figure~\ref{fig-CartoonVersion} and the Appendix), the ambient $t_{\rm cool}/t_{\rm ff}$ ratio is lowest at boundary between those zones.  Therefore, the minimum value of $\min(t_{\rm cool}/t_{\rm ff})$ can depend on how heat input is distributed with radius.  

To see how the structure of a globally balanced system responds to the radial distribution of feedback heating, consider how heating affects evolution of the entropy-profile slope.   The Lagrangian derivative of $\alpha_K(r,t)$ with respect to time is
\begin{equation}
  \frac {d \alpha_K} {dt}  \: = \: \frac {\partial } {\partial t} \left( \frac {\partial \ln K} {\partial \ln r} \right)
  						+ v \cdot \frac {\partial} {\partial r}
						     \left( \frac {\partial \ln K} {\partial \ln r} \right)
 			             \: = \: \frac {\partial } {\partial \ln r} \left( \frac {d \ln K} {d t} \right)
  						+ \left( \frac {\partial \ln K}  {\partial \ln r} \right) 
						     \frac {\partial} {\partial \ln r}
						     \left( \frac {v} {r} \right)
					\; \; ,
  \label{eq-alpha_K-dot1}
\end{equation}
where $v$ is the radial velocity of the Lagrangian shell at $r$.  That radial velocity depends on how quickly the volume bounded by the shell is expanding or contracting due to heating or cooling.  It can be expressed in terms of the volume-weighted average of the change in specific volume ($1/\rho$) within $r$:
\begin{equation}
  \frac {v} {r} = \frac {d \ln r} {dt} = - \frac {1} {3}  \left\langle \frac {d \ln \rho} {dt} \right\rangle 
  \; \; \; \; \; \; , \; \; \; \; \; \; 
  \left\langle \frac {d \ln \rho} {dt} \right\rangle 
  	       \, \equiv  \, \frac {3} {4 \pi r^3} \int_0^r \left(  \frac {d \ln \rho} {dt} \right) 4 \pi r^2 \,  dr  \; \; .
    \label{eq-average_drho-dt}
\end{equation}
The Lagrangian derivative of $\alpha_K$ can therefore be written as
\begin{equation}
  \frac {d \alpha_K} {dt}  = \frac {\partial } {\partial \ln r} \left( \frac {d \ln K} {dt} \right)
  						+ \alpha_K \frac {d \ln \rho} {dt}
						- \alpha_K \left\langle \frac {d \ln \rho} {dt} \right\rangle 
					\; \; .
  \label{eq-alpha_K-dot2}
\end{equation}
In this equation, the first term represents changes in $\alpha_K$ that arise from the radial gradient of the net heating rate.  The second term represents changes that arise from compression or rarefaction of the local entropy gradient.  The third term represents changes that result from expansion or contraction in radius of the Lagrangian shell under consideration.

We are interested in knowing whether $\alpha_K$ increases or decreases with time in response to the distribution of heating and cooling with radius.  If $\alpha_K = 0$, then the answer is relatively simple.  The entropy slope steepens in regions where the net heating rate increases with radius and declines in regions where it decreases with radius.  However, the answer becomes more complicated when $\alpha_K \neq 0$ and the other two terms come into play.  To better understand how the additional terms affect the evolution of the system as a whole, we explore three specific radial distributions of heating and cooling in the following three subsections, and in a fourth subsection (\S \ref{sec-interpretation-of-10}) we bring those explorations to bear on interpreting $\min(t_{\rm cool}/t_{\rm ff}) \approx 10$.

\subsubsection{Two Regions:  Central Heating, Distributed Cooling}
\label{sec-HC}

In the simplest case to consider, heating dominates cooling at the center and cooling dominates heating in the outer regions.  Central heating is the mode of AGN feedback most often implemented in simulations of galaxy evolution.  Examples in the literature are too numerous to mention here, but their generic behavior is captured by the simulation from \citet{Meece_2016arXiv160303674M} in which the kinetic energy of the feedback response is set to zero:  Condensation proceeds until feedback raises the central cooling time of the ambient medium to several~Gyr by creating a large isentropic zone. In such simulations, the amounts of condensed gas and star formation become unacceptably large unless the rise in central cooling time is rapid, which requires a high feedback efficiency.

This outcome occurs because central heating in a globally balanced system corresponds to a radial gradient of $d \ln K / dt$ that is negative, which reduces $\alpha_K$ with time.  The other two terms in equation~(\ref{eq-alpha_K-dot2}) consequently decline in importance.  The result is a central region that is effectively isentropic and also has access to a flow of free energy. These two features are exactly what is needed for condensation to produce multiphase structure in a medium with $t_{\rm cool} \gg t_{\rm ff}$.  In such a globally balanced system, central heating continues to increase the central entropy and can promote condensation until the minimum cooling time in the ambient medium becomes comparable to the age of the universe.  As this happens, the system's ambient medium inevitably evolves toward $\min( t_{\rm cool}/t_{\rm ff} ) \gg 10$.

\subsubsection{Two Regions: Distributed Heating, Central Cooling}
\label{sec-CH}

The opposite case, in which heating exceeds cooling at large radii and cooling exceeds heating at small radii, evolves in the opposite sense.  In such a system, the radial gradient of $d \ln K / dt$ is positive and tends to increase the entropy slope $\alpha_K$.  However, increases in $\alpha_K$ make the second and third terms in equation (\ref{eq-alpha_K-dot2}) more important.  The effects of those other two terms are clearest if heating is insignificant.  Then cooling causes $\rho$ to increase with time in the central cooling-dominated zone on a time scale $\sim t_{\rm cool}$.  It also causes the third term to increase more than the second term, because the average cooling time within a sphere of radius $r$ is generally shorter than the average cooling time at $r$ in a system with a positive entropy gradient.  As a result, the third term limits the entropy gradient to be no greater than the value at which the third term offsets the other two.

One can estimate the limiting value of $\alpha_K$ by assuming that the system remains approximately isothermal as it loses entropy through radiative cooling, which is usually valid for gas confined by a nearly isothermal gravitational potential, as long as $t_{\rm cool} \gg t_{\rm ff}$.  With that assumption, we obtain
\begin{equation}
  \frac {d \alpha_K} {dt}  \, \approx \, \frac {3 \alpha_K} {t_{\rm cool}} 
				\left( \frac {1 - \alpha_K} {2 - \alpha_K} \right) 
      \label{eq-alpha_K-cooling}
\end{equation}
if the volume average in equation~(\ref{eq-average_drho-dt}) is performed assuming $\alpha_K$ is constant with radius.  This estimate shows that cooling produces an inflow that approaches a limiting entropy slope of $\alpha_K \approx 1$, corresponding approximately to $\rho \propto r^{-3/2}$ in gas that remains nearly isothermal, and is in alignment with the self-similar cooling-flow solution derived by \citet{Bertschinger_1989ApJ...340..666B}. In a more realistic cluster-core potential, deviations from isothermality can slightly steepen the entropy slope of a steadily cooling system to $\alpha_K \approx 1.2$, but it always remains close to unity \citep[e.g.,][]{Voit_2011ApJ...740...28V}.
Formally, equation~(\ref{eq-alpha_K-cooling}) also suggests that a cooling region initialized without an entropy gradient remains isentropic, because the cooling time in the isothermal approximation we have adopted is then independent of radius.  In that limit, the effects of temperature gradients and gradients in heat input can no longer be ignored.  However, even if there is some growth of the entropy gradient, it proceeds on a time scale comparable to or longer than the cooling time.  

We conclude that if cooling dominates heating at small radii, then entropy losses and the resulting compression cause the isentropic region to shrink in radius as the center of the system evolves toward a state with $\alpha_K \approx 1$.  Condensation then happens only at small radii, where $t_{\rm cool} \lesssim t_{\rm ff}$.  The result is to focus condensation and mass deposition onto the neighborhood of the black hole \citep[e.g.,][]{LiBryan2012ApJ...747...26L}.  {\em This outcome is highly favorable for coupling of black-hole accretion to the global evolution of the surrounding medium.}  At larger radii, where heating exceeds cooling, the response of $\alpha_K$ depends on the radial gradient of heat input, which depends in turn on the feedback delivery mechanism.  However, the resulting changes in $\alpha_K$ typically proceed on a much longer time scale than the cooling-induced changes to the inner profile, because the heat input required for global balance is spread over a much larger volume, in gas with much lower density and a much longer cooling time.

\subsubsection{Three Regions: Heating-Cooling-Heating}
\label{sec-HCH}

Neither of the simple two-region cases yields a satisfactory explanation for how feedback in a cluster core can maintain a central cooling time $\lesssim 1$~Gyr along with $\min(t_{\rm cool} / t_{\rm ff}) \approx 10$ for periods of several~Gyr.  We must therefore take one more small step toward complexity and consider what happens if there are {\em three} nested regions in which the sign of $d \ln K / dt$ alternates, with heating exceeding cooling at the smallest and largest distances from the black hole, while cooling exceeds heating at intermediate radii.

\begin{figure*}[t]
\begin{center}
\includegraphics[width=4.7in, trim = 0.0in 0.0in 0.0in 0.0in]{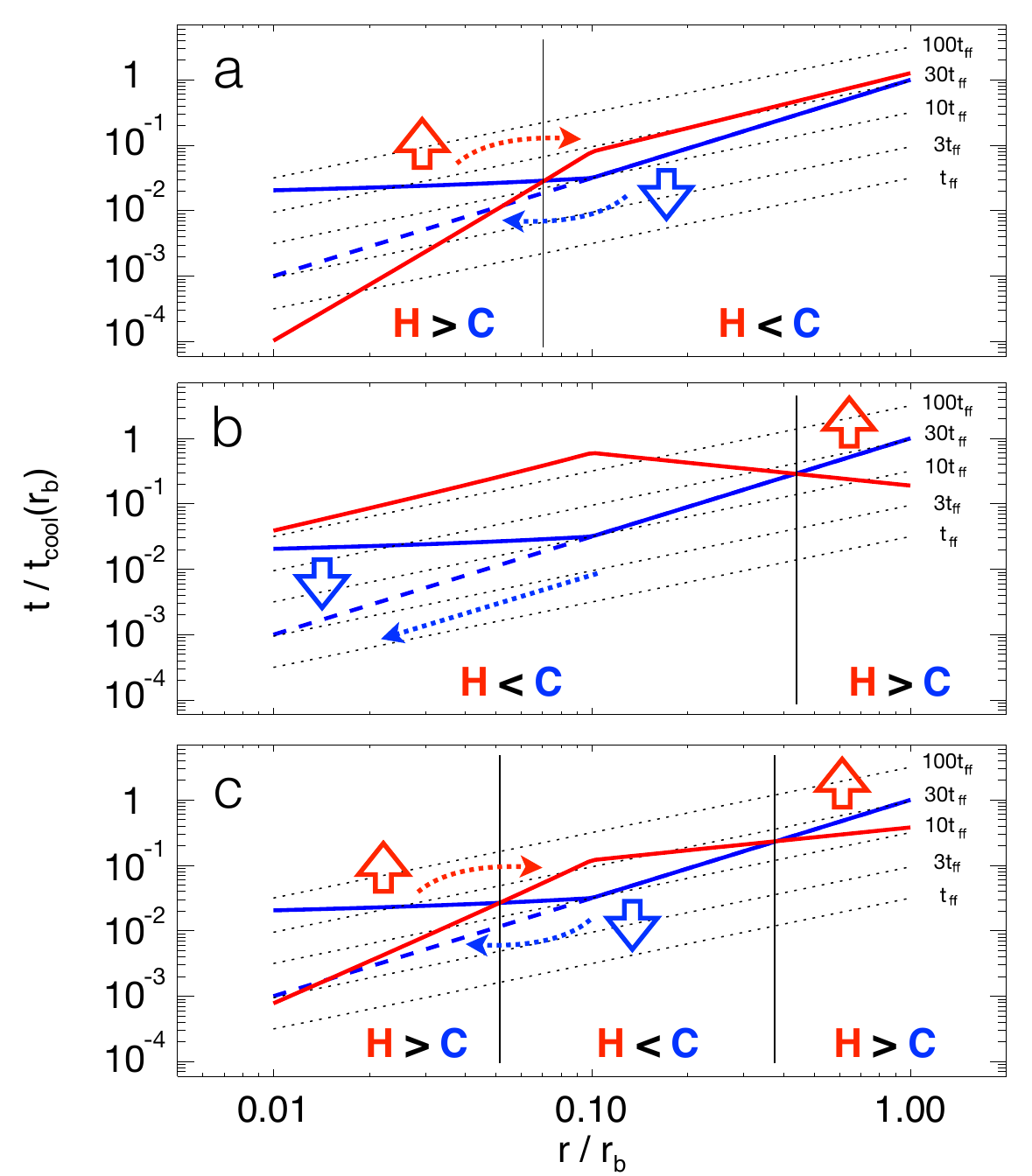} \\
\end{center}
\caption{ \footnotesize 
Schematic illustrations of configuration changes in globally balanced systems.  In each panel cooling equals heating within a boundary radius $r_{\rm b}$, and solid blue lines show identical cooling-time profiles derived from the toy model in the Appendix with $r_0 = 0.1 r_{\rm b}$ and $\alpha_K = 1.0$. The blue dashed line shows the cooling-time profile that would result if the power-law entropy profile extended to small radii.  Dotted black lines show the slope of $t_{\rm ff}(r)$. The solid red line in each panel shows the local heating timescale $t_{\rm heat}$, corresponding to heat input that has a power-law dependence on radius  It is normalized so that the total heat input $\dot{E}(r_{\rm b})$ equals the total luminosity generated by radiative cooling within $r_{\rm b}$.  Panel (a) shows $t_{\rm heat}$ for a heat-input distribution ($\dot{E}\propto r^{0.3}$) for which heating per unit volume exceeds cooling at small radii, causing the central entropy to rise with time, and falls short of cooling at large radii, (see \S \ref{sec-HC}).  Panel (b) shows a heat-input distribution ($\dot{E} \propto r^2$) that fails to match cooling at small radii but exceeds it at large radii, causing the center to approach a steady cooling flow with $\alpha_K = 1$ (see \S \ref{sec-CH}).  Panel (c) shows a heat-input distribution ($\dot{E} \propto r$) in which heating exceeds cooling at small radii, falls short at intermediate radii, and exceeds it again at large radii (see \S \ref{sec-HCH}).   
\vspace*{1em}
\label{fig-GlobalBalance}}
\end{figure*}

Figure~\ref{fig-GlobalBalance} schematically shows how the response to feedback of this three-region system contrasts with the two-region cases we have just discussed.  The radial decline in $d \ln K /dt$ across the boundary between the innermost two regions causes convection that drives the inner part of the system toward an isentropic configuration that is prone to condensation.  Meanwhile, the rise in $d \ln K /dt$ across across the boundary between the outermost two regions maintains a large-scale entropy gradient that suppresses thermal instability through buoyancy damping.  The natural outcome of such a feedback response is a long-lasting configuration with an inner isentropic zone and an outer power-law zone (see Figure~\ref{fig-CartoonVersion} again).  In this configuration central production of cold clouds can fuel a feedback response that prevents condensation in the outer zone.  The Appendix presents a toy model that makes this conceptual argument more quantitative.

\subsubsection{Interpretation of $t_{\rm cool} / t_{\rm ff} \approx 10$}
\label{sec-interpretation-of-10}

This chain of reasoning leads to an interpretation of $\min (t_{\rm cool} / t_{\rm ff}) \approx 10$ with several facets.  Section \ref{sec-HCH} outlines the radial distribution of feedback heat input needed to keep $\min (t_{\rm cool} / t_{\rm ff})$ steady but does not explain how this balance remains stable.  If the system is to remain near a particular balance point for time periods much longer than the cooling time, then fluctuations in $\min (t_{\rm cool} / t_{\rm ff})$ need to produce compensating responses.  Here we suggest that volatility places a lower bound on $\min (t_{\rm cool} / t_{\rm ff})$ and that inefficient central heating places an upper limit on $\min (t_{\rm cool} / t_{\rm ff})$.

First, consider how such a system responds to a decline in $\min (t_{\rm cool} / t_{\rm ff})$.   As the value of this ratio drops below $\approx 10$, the system becomes increasingly vulnerable to feedback outbursts fueled by condensation.  Outbursts that uplift gas at speeds close to the circular velocity of the gravitational potential promote condensation of gas that initially has $t_{\rm cool} / t_{\rm ff} \lesssim 10$ (see \S \ref{sec-tc_isoK}).  Condensation then removes the lowest entropy gas from the ambient medium, leaving behind higher entropy gas with a longer cooling time.  Promoting condensation with uplift becomes progressively more difficult as this process continues, because $\min (t_{\rm cool} / t_{\rm ff})$ rises and uplift of large amounts of gas at speeds substantially exceeding the circular velocity is unlikely.  Therefore, this part of the feedback loop restores the system to $\min (t_{\rm cool} / t_{\rm ff}) \approx 10$.

Next, consider the response to a rise in $\min (t_{\rm cool} / t_{\rm ff})$ driven by central heating.  Condensation can continue in the isentropic zone but is spread over a larger region, making it more difficult for condensed gas clouds to shed enough angular momentum to accrete onto the black hole. In nature, this reduction in the fuel supply might be all that is needed to diminish central heating and restore $\min (t_{\rm cool} / t_{\rm ff}) \approx 10$.  In simulations, however, spatial resolution is more limited and subgrid feedback algorithms have much less of an angular momentum problem to overcome than in nature.  Nevertheless, simulations with precipitation-fueled bipolar outflows also manage to self-regulate.

We suggest that in these simulations, and maybe also real galactic systems, the restoring response to a rise in $\min (t_{\rm cool} / t_{\rm ff})$ comes about because the central region less efficiently taps the feedback energy that is transported through it.  The Appendix presents this suggestion more quantitively in the context of our toy model.  Global regulation requires the total amount of feedback heating to be at least as great as losses to radiative cooling, and we have shown that much of that feedback energy must pass through the isentropic zone without thermalizing there.  A rise in $\min (t_{\rm cool} / t_{\rm ff})$ corresponds to a decrease in the density of the isentropic zone, and if $\alpha_K > 2/3$, also a decrease in its column density.  It is therefore plausible that reducing the density of the isentropic zone reduces the fraction $f_0$ of feedback energy deposited there \citep[see, for example, the recent simulations by][]{TB_jets_2016MNRAS.tmpL..44T}.  

To appreciate the consequences of an inverse relationship between $f_0$ and $K_0$, consider first what happens if $f_0$ remains constant.  Suppose the system begins in a volatile state, with $\min (t_{\rm cool} / t_{\rm ff}) < 10$, so that condensation feeds the black-hole engine and boosts the output rate $\dot{E}$ of feedback power.  Increasing condensation then boosts $\dot{E}$ until $f_0 \dot{E}$ balances the radiative luminosity of an isentropic zone with $\min (t_{\rm cool} / t_{\rm ff}) \approx 10$.  As $\min (t_{\rm cool} / t_{\rm ff})$ rises beyond that point, condensation should diminish, but a large cold-gas reservoir may already have accumulated, which can continue to fuel the outflow.  Upward fluctuations in $\dot{E}$ fueled by that cold-gas reservoir can in principle cause $K_0$ and $\min (t_{\rm cool} / t_{\rm ff})$ to grow further.  However, if $d f_0 /d K_0 < 0$, then increases in $K_0$ diminish the heat supply to the isentropic zone and allow cooling to return the system to its original value of $K_0$.  Simultaneous long-term regulation of both the isentropic zone and the power-law zone can therefore be achieved if radiative cooling of the power-law zone is $\lesssim (1 - f_0) \dot{E}$.  Precise balance between heating and cooling is not necessary in the power-law zone because the timescale for configuration changes at large radii is long, and feedback regulation of the isentropic zone will quickly respond to slow feedback-driven expansion of the power-law zone.

\section{Illustrations from Simulations}
\label{sec-Illustrations}

Our objective until this point in the paper has been to understand the physical principles that govern precipitation-regulated feedback, both in nature and in simulations of galaxy evolution.  Now it is time to apply those principles to interpret what happens in such simulations, using an idealized cluster-core simulation from \citet{Li_2015ApJ...811...73L} as a paradigmatic example.  In that simulation, condensation produces cold gas clouds that feed a central black hole, which responds by producing a bipolar outflow.  This mode of feedback regulates the system so that $\min( t_{\rm cool} / t_{\rm ff}) $ remains in the range $5 \lesssim \min( t_{\rm cool} / t_{\rm ff}) \lesssim 20$ for long time periods while the outflows are active.  However, star formation eventually consumes the cold gas that the black hole needs for fuel.  When that cold-gas reservoir becomes depleted, the outflow shuts down, cooling is now unopposed, and $\min( t_{\rm cool} / t_{\rm ff})$ declines until condensation renews the supply of cold gas.  The outflow then resumes, restoring regulation at $5 \lesssim \min( t_{\rm cool} / t_{\rm ff}) \lesssim 20$.

\begin{figure*}[t]
\begin{center}
\includegraphics[width=6.0in, trim = 0.0in 0.0in 0.0in -0.1in]{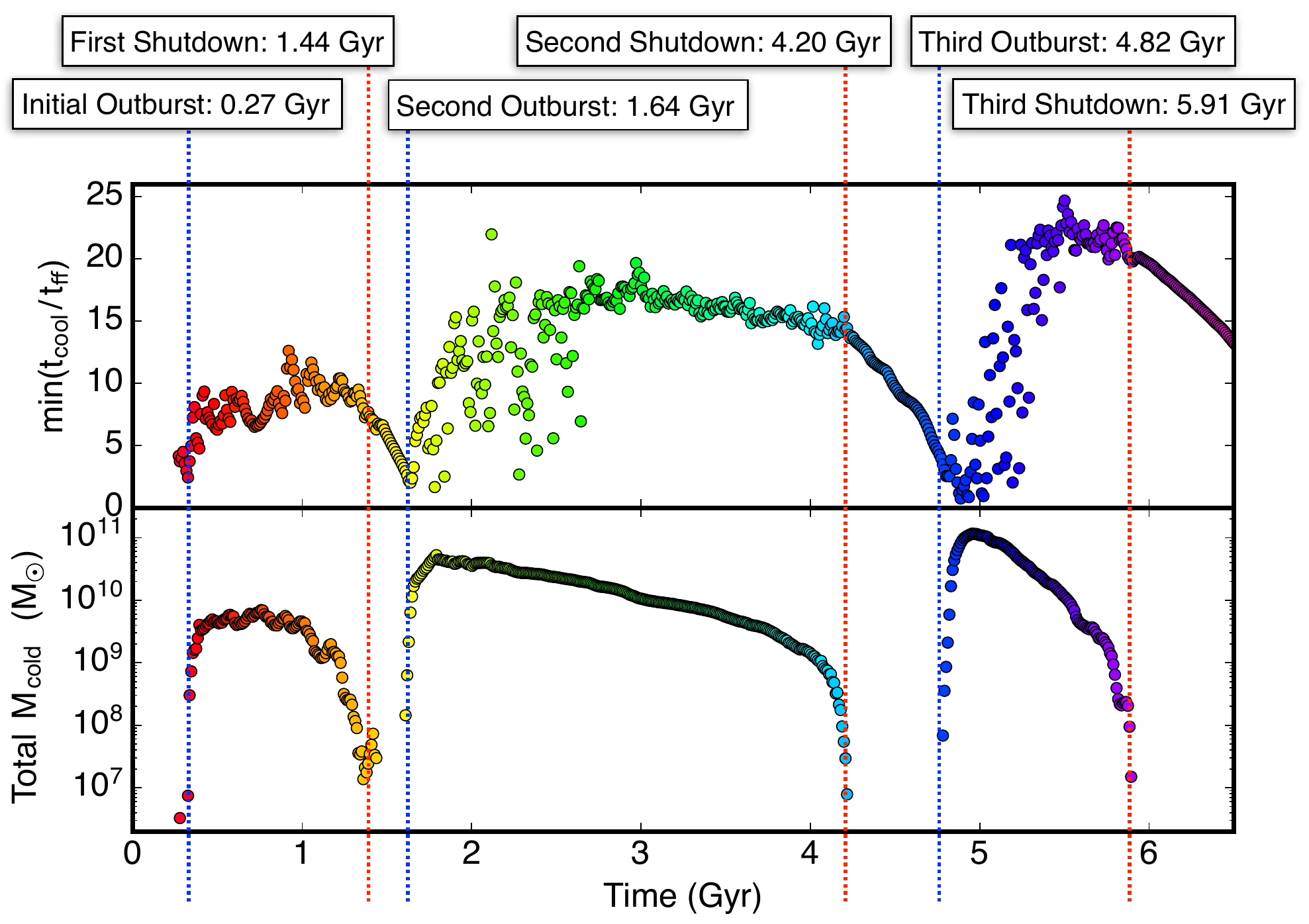} \\
\end{center}
\caption{ \footnotesize 
Outburst history of precipitation-regulated AGN feedback in a simulation from \citet{Li_2015ApJ...811...73L}.  The top panel shows the minimum value of $t_{\rm cool} / t_{\rm ff}$ in the ambient medium at each moment in time, and the bottom panel shows the amount of cold gas available for accretion.  Cold gas rapidly accumulates at the onset of each outburst, but the feedback response regulates the condensation rate as $\min(t_{\rm cool} / t_{\rm ff})$ rises.   Feedback continues to regulate condensation while cold gas is plentiful, but shuts down as star formation uses up the cold gas supply.  The value of $\min(t_{\rm cool} / t_{\rm ff})$ subsequently declines until cooling triggers another feedback outburst.
\vspace*{1em}
\label{fig-OutburstHistory}}
\end{figure*}

The \citet{Li_2015ApJ...811...73L} simulation is far from a complete treatment of all the relevant physics.  Perhaps most importantly, it is not subject to cosmological structure formation, and is therefore artificially symmetric and undisturbed.  It is quite likely that the cold-gas reservoirs in real systems of this type are never fully depleted, in which case there would be no time periods during which the central cooling time would need to decline below $\min( t_{\rm cool} / t_{\rm ff}) \sim 5$ in order to re-initiate condensation.  However, those episodes of cold-gas depletion in the \citet{Li_2015ApJ...811...73L} simulation turn out to be extremely useful for illustrating the respective roles of unopposed cooling, uplift-driven condensation, buoyancy damping, and feedback regulation, as this section will demonstrate.

\subsection{Simulated Feedback Cycles}

Figure~\ref{fig-OutburstHistory} shows the feedback cycles in the \citet{Li_2015ApJ...811...73L} simulation. At the beginning, the gas is homogeneous at each radius and remains homogenous as it cools, until condensation triggers strong feedback at $t = 0.27$~Gyr.  The amount of gas that is present within the central 20~kpc prior to the feedback outburst can be estimated by noting that $\alpha_K \approx 1$ and $K \approx 10 \, {\rm keV \, cm^2}$ at 20~kpc, giving a gas-mass estimate $\sim 3 \times 10^{11} \, M_\odot$ for a gas temperature $\sim 3$~keV.   Approximately $\sim 4 \times 10^9 \, M_\odot$ of this gas condenses within $\sim 200$~Myr and triggers feedback that leaves the system in a steady feedback-regulated state with $\min( t_{\rm cool} / t_{\rm ff}) \approx 10$ for the next $\sim 1$~Gyr, until star formation uses up all the cold gas.

Two more outburst and shutdown cycles follow the first one, and each cycle proceeds similarly.  Without heat input, the core cools homogeneously and develops a power-law central entropy gradient that focuses condensation on the black hole.  When the black hole erupts, uplift stimulates rapid condensation and spreads condensing gas over a much larger volume.  After each eruption, the system is left in a long-lasting, well-regulated state with a flat central entropy gradient.

\subsection{Anatomy of the Specific Entropy Distribution}

\begin{figure*}[t]
\begin{center}
\includegraphics[width=5.5in, trim = 0.0in 0.0in 0.0in 0.0in]{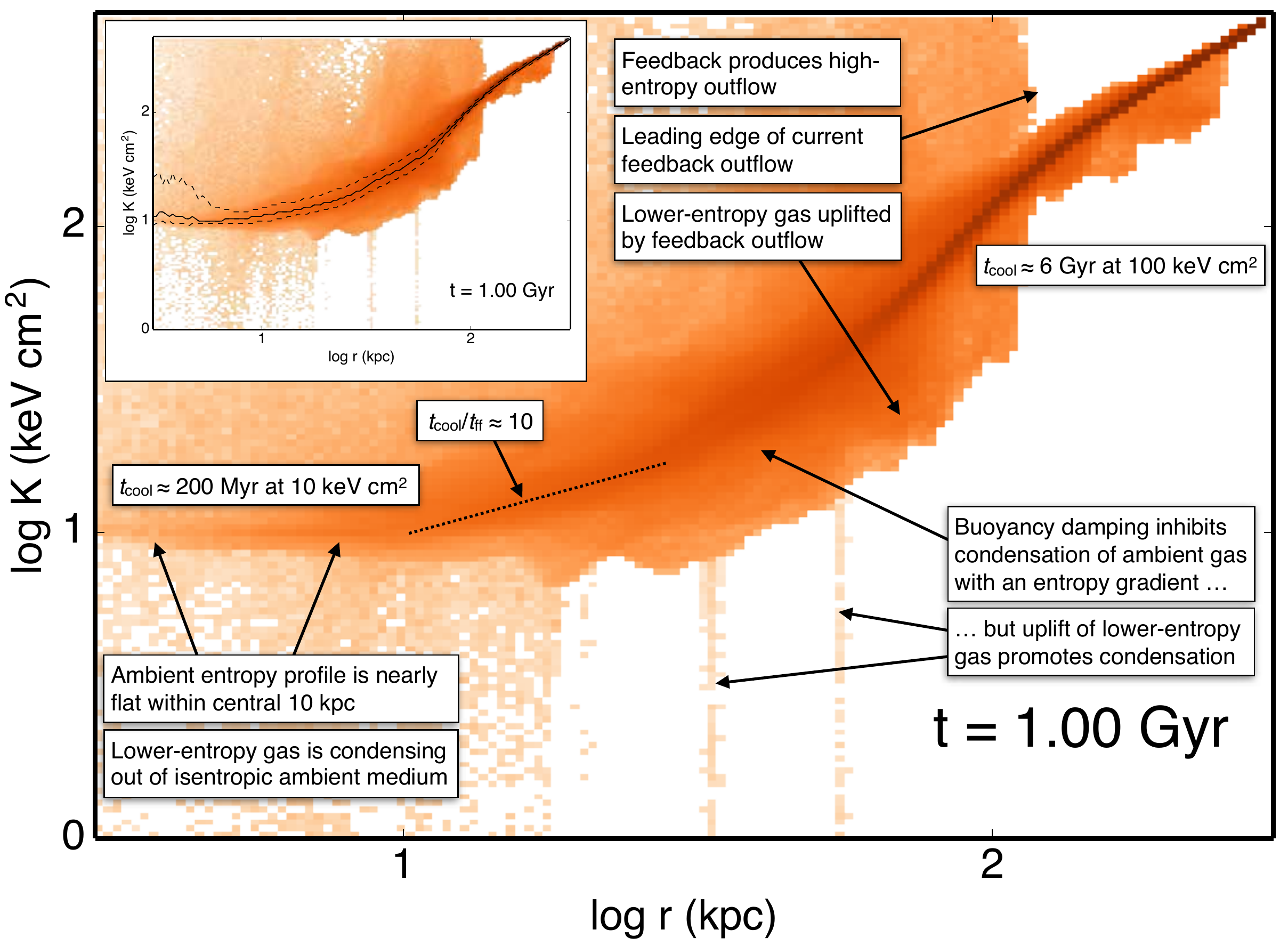} \\
\end{center}
\caption{ \footnotesize 
Anatomy of the specific entropy distribution at $t = 1$~Gyr in the simulation from Figure~\ref{fig-OutburstHistory}.  Shades of orange show the relative amounts of gas in the uncondensed component with specific entropy $K$ at radius $r$.  The median entropy value (shown with a solid black line in the inset figure) is nearly constant with radius inside of 10~kpc and steadily rises at larger radii.  The dispersion in entropy (dashed lines in the inset figure show the 20\%-80\% percentile range) declines with increasing radius.  Buoyancy damping largely suppresses condensation where the entropy gradient is significant but cannot suppress condensation where the entropy gradient is shallow.  The transition between these two regions is where $t_{\rm cool} / t_{\rm ff} \approx 10$ (shown with a dotted line).
\vspace*{1em}
\label{fig-Anatomy}}
\end{figure*}

In order to understand why these cycles unfold as they do, it is important to look closely at the distribution of specific entropy $K$ at each radius.  Figure~\ref{fig-Anatomy} shows the distribution of $K$ as a function of $r$ at $t = 1$~Gyr, which is in the midst of the first feedback cycle.  Broadly speaking, the entropy profile is approximately isentropic in the inner regions and approximately a power law with $\alpha_K \approx 1$ in the outer regions.  The transition from $\alpha_K \ll 1$ to $\alpha_K \approx 1$ is at the radius where $t_{\rm cool} / t_{\rm ff} \approx 10$.  However, gas at any given radius can span a wide range in $K$.

The actual dispersion in specific entropy is not quite as broad as the figure's logarithmic heat map makes it seem.  Lines in the inset figure show the median entropy at each radius (solid line) and the 20\%-80\% percentile range (dashed lines).  Despite the wide range of extremes in entropy, the dispersion is actually quite narrow at large radii.  The extremes are associated with the high-entropy outflow and the lower-entropy gas uplifted by it \citep{LiBryan2014ApJ...789..153L}, but buoyancy damping keeps most of the gas nearly homogeneous at radii where the entropy gradient is significant.

In the center, where the entropy gradient vanishes, buoyancy damping is less successful at suppressing entropy perturbations, and thermal instability is able to produce condensation (\S \ref{sec-GeneralConsiderations}).  Gas blobs below the median entropy in this zone are declining in specific entropy and will soon join the condensed component.  Outside of this thermally unstable zone, condensation is scarcer.  At the epoch shown, there are two isolated patches of multiphase gas at larger radii, roughly 30~kpc and 50~kpc from the center.  Buoyancy damping cannot suppress condensation in these patches because they have been uplifted by the outflow (\S \ref{sec-tc_isoK}) and are on nearly ballistic trajectories.

\subsection{Initial Outburst, Uplift, and Condensation}

\begin{figure*}[t]
\includegraphics[width=7.0in, trim = 0.0in 0.0in 0.0in 0.0in]{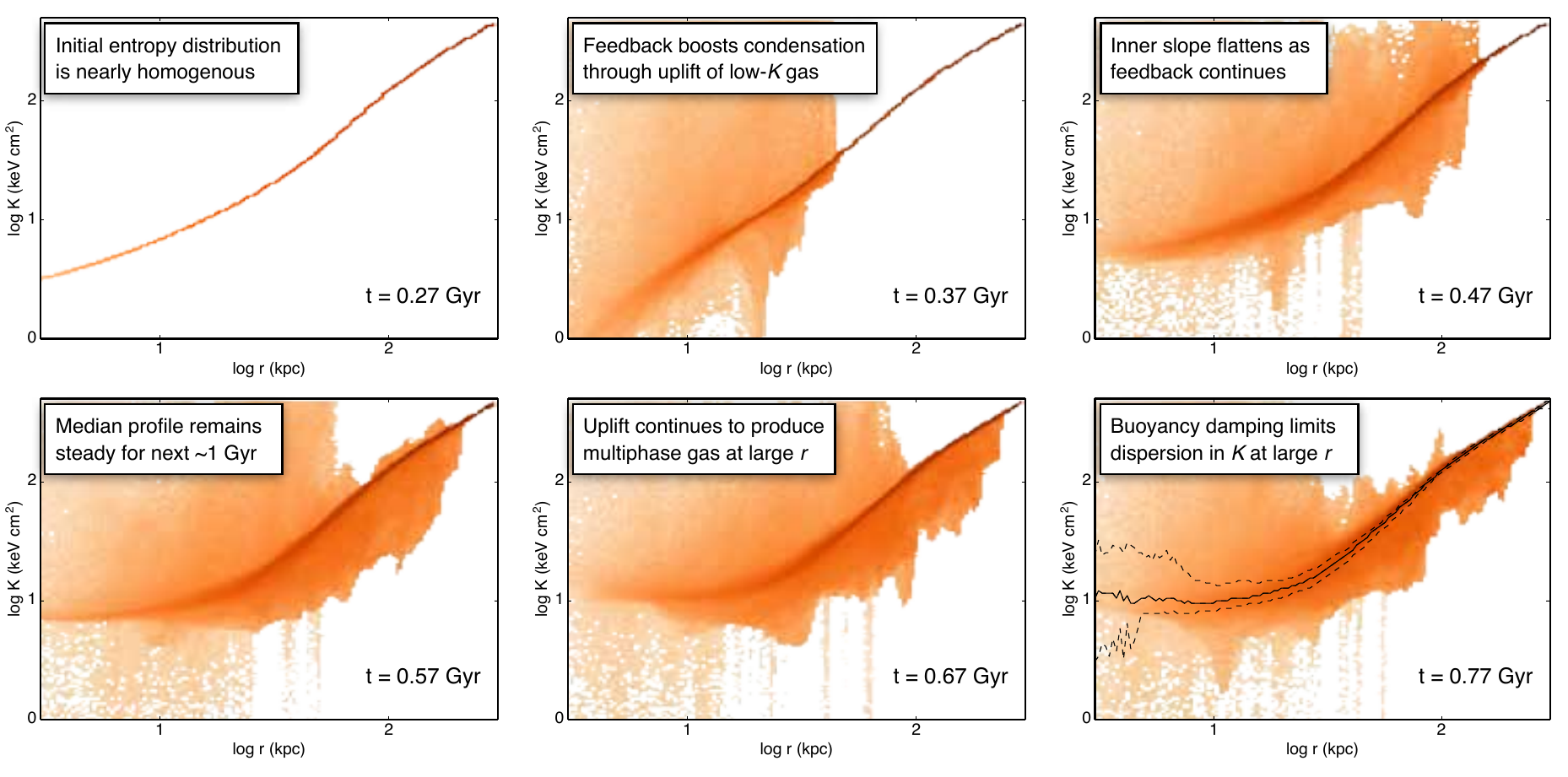} \\
\caption{ \footnotesize 
Evolution of the specific entropy distribution in the simulation from Figure~\ref{fig-OutburstHistory} during triggering of the initial episode of feedback. The outburst of kinetic energy from the center begins when the central entropy becomes low enough for condensation.  At that moment, the median entropy distribution is nearly a power law with $K \stackrel {\propto} {\sim} r$, and $\min(t_{\rm cool} / t_{\rm ff}) \sim 2$.  Shortly thereafter, the feedback outburst flattens the median entropy profile within $\sim 10$~kpc and stabilizes at $\min(t_{\rm cool} / t_{\rm ff}) \sim 10$ for the next $\approx 1$~Gyr.
\vspace*{5em}
\label{fig-InitialOutburst}}
\end{figure*}

The six panels of Figure~\ref{fig-InitialOutburst} show how the entropy distribution evolves during the first outburst of feedback.  Before the outburst erupts at $t = 0.27$~Gyr, the $K$ distribution at each $r$ is still nearly homogenous for two reasons: (1) pure cooling without compensating heating does not lead to phase separation (\S \ref{sec-unmixing}), and (2) even if there were some entropy perturbations, they would be damped by buoyancy (\S \ref{sec-BuoyancyDamping}).  The next panel ($t = 0.37$~Gyr) shows the system 100~Myr after the beginning of the outburst, and the leading edge of the high-entropy outflow (at $\sim 40$~kpc) is well defined.  Gas at larger radii is still undisturbed, and the median entropy still follows an approximate power law to $\ll 10$~kpc.  However, there are significant amounts of condensing low-entropy gas at $\sim 20$~kpc.  That cannot be gas condensing out of the ambient medium at 20~kpc, because the median cooling time there was $\gtrsim 200$~Myr at the time the outburst began.  Instead, it is low-entropy gas that was transported outward from smaller radii by the outflow, and uplift is allowing it to condense.

After another 100~Myr (at $t = 0.47$~Gyr), an isentropic central region has started to develop at $K_0 \approx 10 \, {\rm keV \, cm^2}$, where the cooling time is $\approx 200$~Myr.  In order for an isentropic central region to form, the bulk of the lower-entropy gas must be eliminated through either condensation, heating, or mixing.  Figure~\ref{fig-OutburstHistory} shows that less than $\sim 10^{10} \, M_\odot$ condenses during this event, implying that the rest of the $\sim 3 \times 10^{11} \, M_\odot$ initially within the central $\lesssim 20$~kpc is either heated or mixed into gas of greater entropy.

The remaining three frames, spanning the next 300~Myr, show that the system becomes well-regulated after arriving at a state with an isentropic core and a power-law outer region (see \S \ref{sec-HCH}).  Condensation continues in the isentropic core at $\lesssim 20$~kpc but proceeds only in isolated patches beyond where the entropy gradient steepens, and those patches are moving on approximately ballistic trajectories.  The outflow continues to populate the extremes of the entropy distribution.  However, buoyancy damping causes those extremes to evolve toward the median, as shown by the lines in the lower-right panel, which mark the median and 20\%-80\% percentile range, as in the Figure~\ref{fig-Anatomy} inset.

\subsection{Second Outburst}

\begin{figure*}[t]
\includegraphics[width=7.0in, trim = 0.0in 0.0in 0.0in 0.0in]{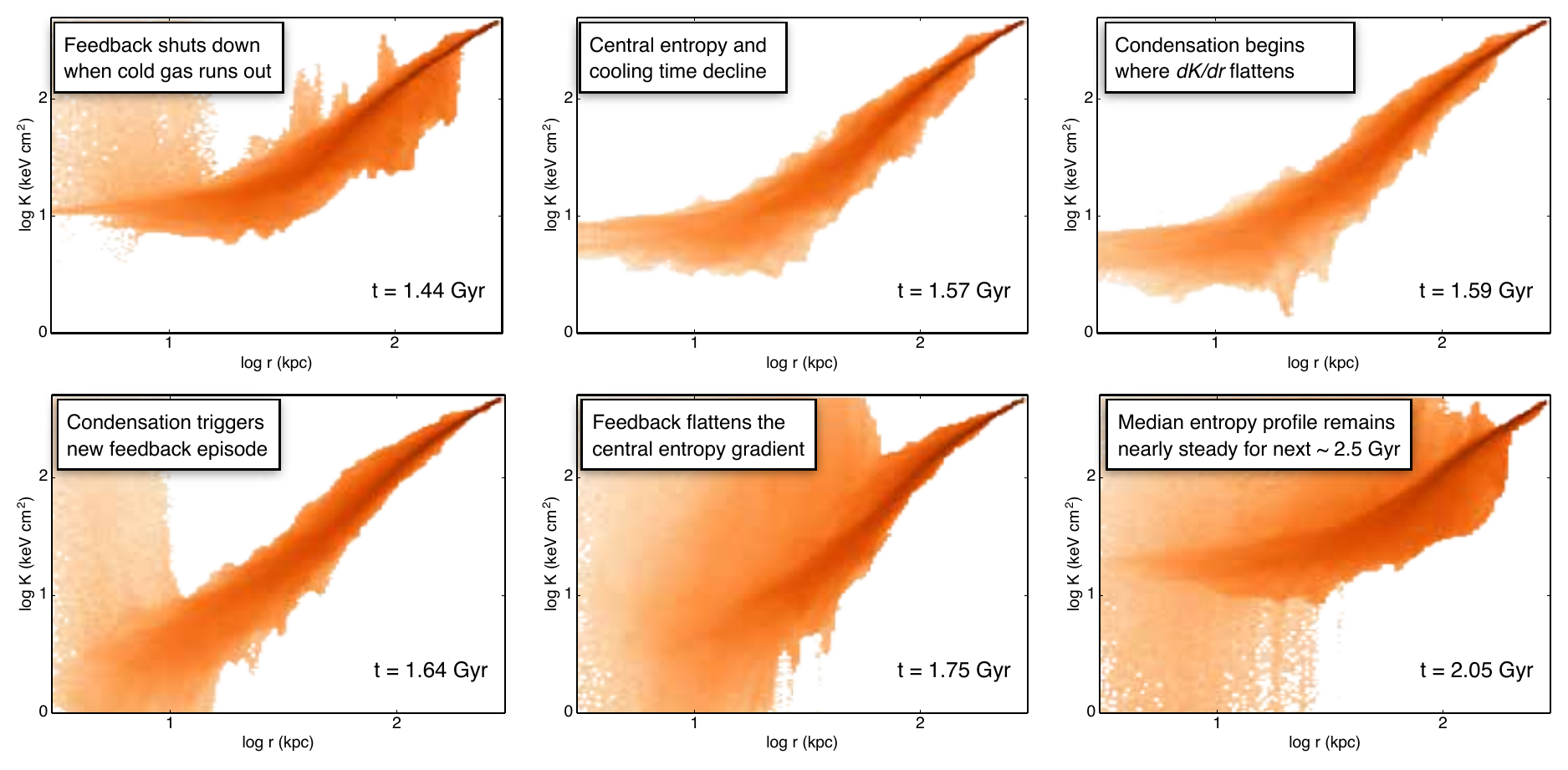} \\
\caption{ \footnotesize 
Evolution of the specific entropy distribution in the simulation from Figure~\ref{fig-OutburstHistory} during triggering of the second episode of feedback. When the supply of cold gas for accretion runs out, the central cooling time is $\approx 200$~Myr.  The central gas then cools uniformly, because feedback is not supplying the flow of free energy needed to promote phase separation.  Meanwhile, buoyancy damping is reducing the entropy perturbations at large radii.  Condensation therefore begins at the outer edge of the isentropic region, where $t_{\rm cool} / t_{\rm ff}$ is smallest and buoyancy damping cannot operate.
\vspace*{1em}
\label{fig-SecondOutburst}}
\end{figure*}

Figure \ref{fig-SecondOutburst} shows six more frames, this time spanning the period from the end of the first feedback episode to a time approximately 400~Myr after the beginning of the second.  At $t = 1.44$~Gyr, the cold fuel has just run out, causing the AGN to shut down, but the residual outflow has not finished propagating.  Some of the gas is far from the median entropy level, and the central cooling time is $\approx 200$~Myr.  During the next $\sim 100$~Myr, buoyancy damping reduces the extremes in entropy at each radius in the power-law zone.  Meanwhile, the central cooling time begins to decline, but it does so almost uniformly, because there is no source of free energy to cause phase separation (\S \ref{sec-unmixing}). Condensation finally starts to happen as the simulation approaches $t = 1.59$~Gyr, and it develops most quickly at the boundary between the isentropic and power-law zones, because that is where $t_{\rm cool}/t_{\rm ff}$ is smallest (\S\ref{sec-GlobalThermalBalance}).

Once condensation begins, cooling gas quickly converges toward the center and triggers a second feedback outburst at $t = 1.64$~Gyr.  At this moment, the median entropy distribution is nearly a  power law with $\alpha_K \approx 1.3$ from 100~kpc to within 10~kpc.  This median slope is almost identical to the triggering slope at $t = 0.37$~Gyr in Figure~\ref{fig-InitialOutburst}, but the pre-triggering entropy distribution at $\lesssim 10$~kpc is much less homogeneous because of the condensation that started happening 50~Myr earlier.

A major consequence of this pre-existing inhomogeneity is that uplift induces condensation far more effectively when the outflow begins, yielding $\sim 5 \times 10^{10} \, M_\odot$ of cold gas during the next 200~Myr (see Figure~\ref{fig-OutburstHistory}).  The gas mass within $\sim 20$~kpc of the center as condensation begins is similar to the $\sim 3 \times 10^{11} \, M_\odot$ that was there during the first outburst, but 10 times more of it condenses (see Figure~\ref{fig-OutburstHistory}).  The rest must be either heated or mixed to $\gtrsim 20 \, {\rm keV \, cm^2}$ because the central region becomes nearly isentropic at this level by $t = 2.05$~Gyr, as shown in the final panel of Figure~\ref{fig-SecondOutburst}.

After the second outburst produces another reservoir of cold-gas fuel, the system enters a long period of steady self-regulation.  The median entropy profile within $\sim 20$~kpc stays flat at $K_0 \approx 20 \, {\rm keV \, cm^2}$, so that $\min (t_{\rm cool} / t_{\rm ff}) \approx 15$, apart from some stochastic excursions as the black-hole energy output fluctuates.  However, the cold fuel eventually runs out after sustaining $\approx 2.5$~Gyr of activity, setting the stage for the third outburst.

\subsection{Third Outburst}

\begin{figure*}[t]
\includegraphics[width=7.0in, trim = 0.0in 0.0in 0.0in 0.0in]{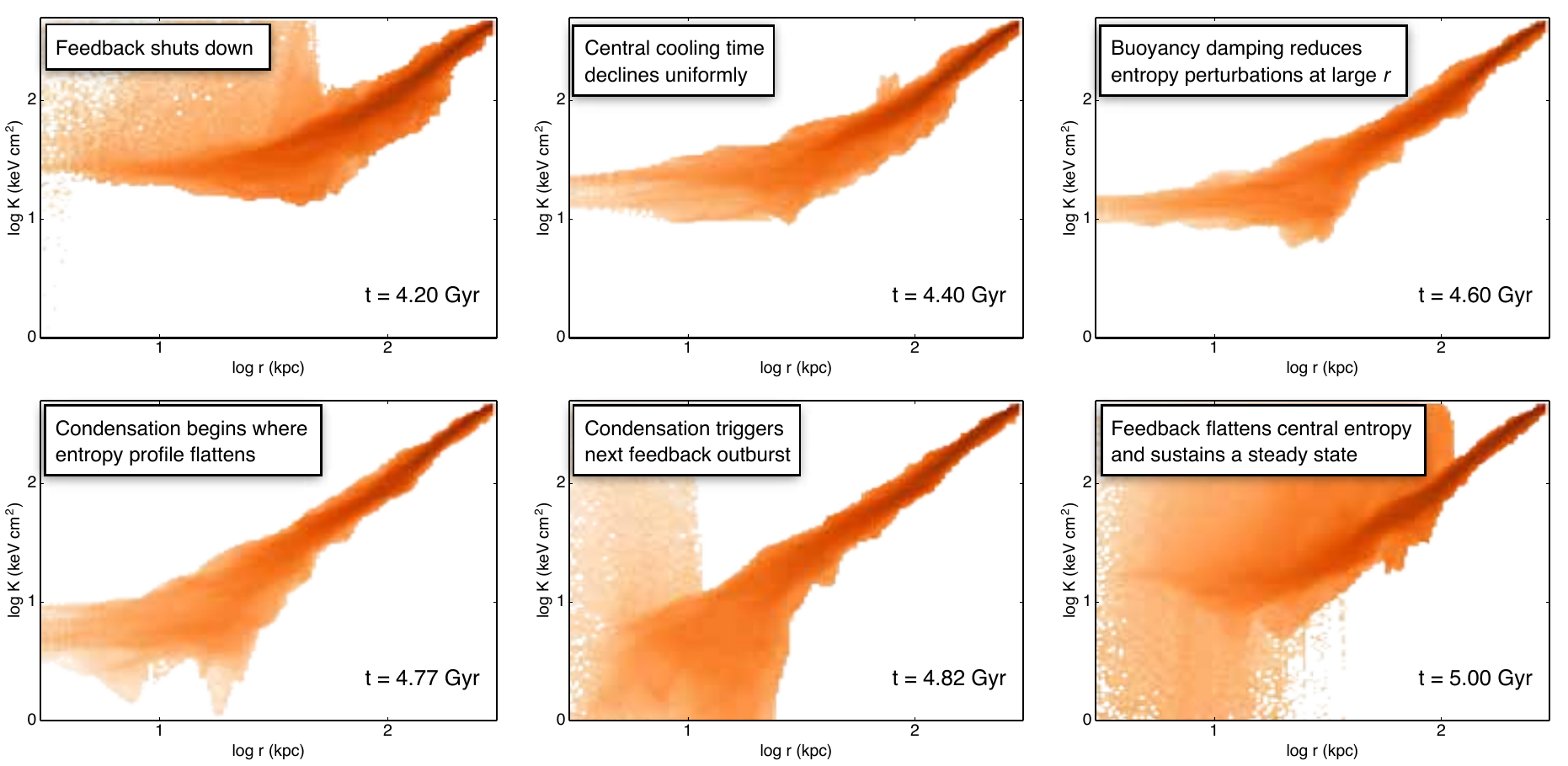} \\
\caption{ \footnotesize 
Evolution of the specific entropy distribution in the simulation from Figure~\ref{fig-OutburstHistory} during triggering of the third episode of feedback. This triggering episode proceeds similarly to the second one in Figure~\ref{fig-SecondOutburst}, except that the quiescent period is longer because central cooling time is $\approx 0.6$~Gyr when the supply of cold gas runs out.  
\vspace*{4em}
\label{fig-ThirdOutburst}}
\end{figure*}

Figure~\ref{fig-ThirdOutburst} concentrates on the events leading up to the third outburst, which require more time to unfold because the central cooling time of the system is $\approx 600$~Myr when the cold fuel runs out.  Complete depletion of cold fuel could be a consequence of the idealizations in this simulation that never happens in real cluster cores.  We mention this possibility because very few real clusters have central cooling times $\lesssim 1$~Gyr and no multiphase gas \citep{VoitDonahue2015ApJ...799L...1V}.  However, the period of quiescence before the third outburst in this simulation beautifully illustrates the physics of AGN triggering via condensation, and similar processes are likely to operate in less idealized systems.

The events that lead to triggering are almost identical to those in Figure~\ref{fig-SecondOutburst}, but here the progression of thermal instability is easier to see.  Buoyancy damping again reduces entropy perturbations in the power-law zone, and the central regions again cool uniformly because there is no source of free energy to promote phase separation.   However, buoyancy damping is least effective at the junction between the isentropic and power-law zones, where $t_{\rm cool}/t_{\rm ff}$ is smallest and is also declining with time.  That junction moves inward as the isentropic zone loses thermal energy and becomes more highly compressed.  As a result, the median entropy profile approaches a pure power law in radius, while thermal instability progresses toward condensation at the outer edge of the isentropic zone.  Condensation then occurs between $t = 4.77$~Gyr and $t = 4.82$~Gyr.  It happens first at $r \approx 20$~kpc, but soon the entire region within 20~kpc contains multiphase gas.

The feedback outflow then sweeps through the condensing region, and 200~Myr later ($t = 5.00$~Gyr) a new isentropic zone has begun to form.  This triggering period produces even more condensation than the previous two outbursts, yielding $\approx 10^{11} \, M_\odot$ of cold gas.  With this fuel source, the feedback outflow stabilizes the system's global configuration at $K_0 \approx 30 \, {\rm keV \, cm^2}$ and $\min (t_{\rm cool} / t_{\rm ff}) \approx 20$ for another $\sim 1$~Gyr.

\subsection{Not-Quite-Ballistic Multiphase Regions}

During the long periods of relatively steady feedback regulation between these triggering events, condensation happens mostly within the isentropic zone, where it is not inhibited by buoyancy damping.  However, multiphase regions are sometimes observed far outside the isentropic zone, at radii where $\alpha_K \approx 1$ and $t_{\rm cool}/t_{\rm ff} > 10$ in the ambient medium, implying that buoyancy damping should be suppressing condensation.  In plots of the $K(r)$ distribution, those multiphase regions show up as isolated vertical stripes extending far below the median entropy level.

\begin{figure*}[t]
\includegraphics[width=7.0in, trim = 0.0in 0.0in 0.0in 0.0in]{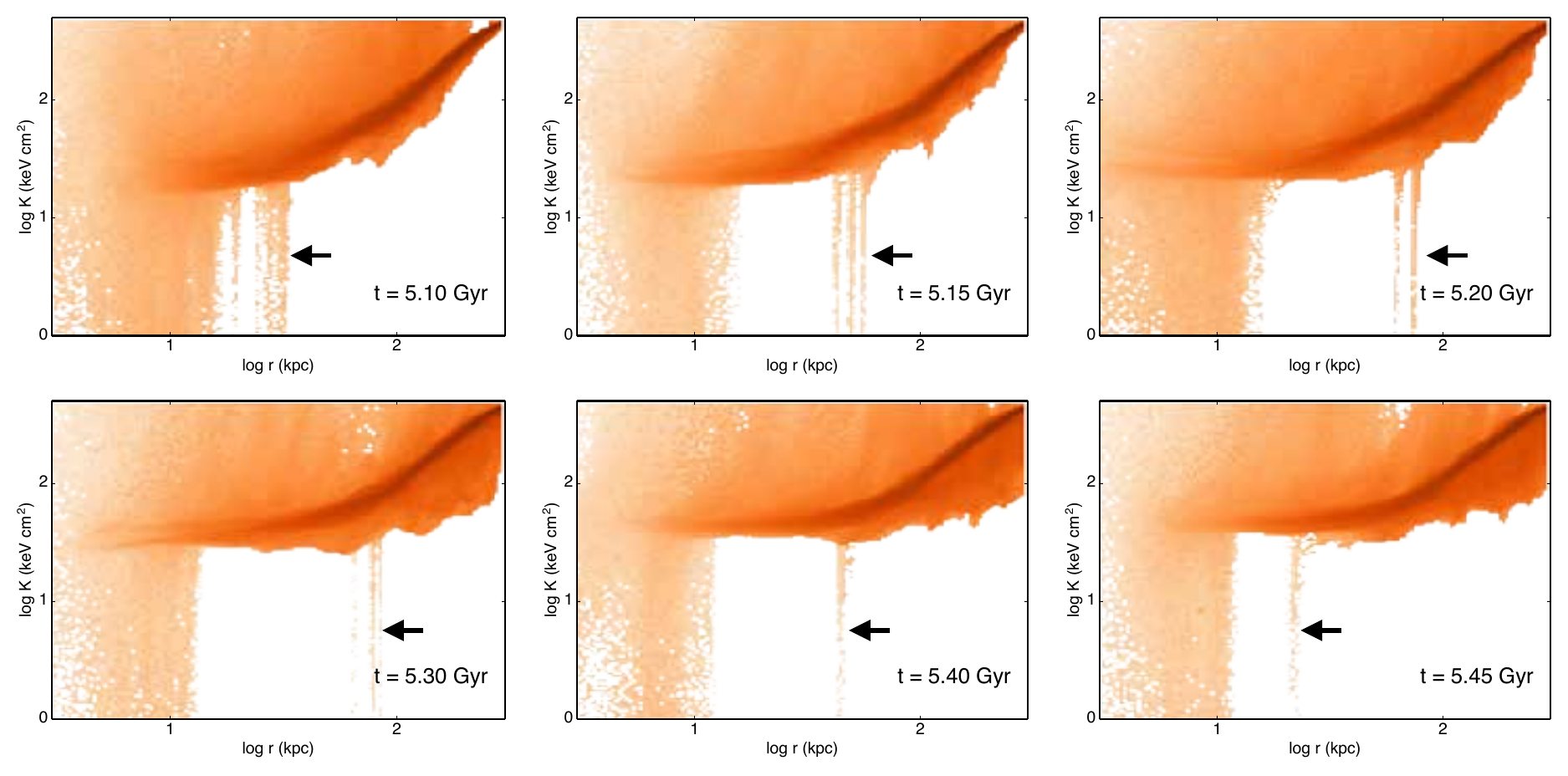} \\
\caption{ \footnotesize 
Isolated region of multiphase gas rising and falling through the power-law zone.  At large radii, isolated multiphase regions appear as conspicuous stripes extending directly downward in $K$ far below the median entropy.  They remain multiphase as they move vertically within the simulation environment and sometimes develop spatial substructure, which manifests as multiple components in the radial domain.  Because they are long-lasting, their vertical trajectories can be tracked as the simulation progresses.  The black arrows in these panels show how a particular multiphase region is moving with time.  Its trajectory rises and falls at a peak speed of $\approx 500 \, {\rm km \, s^{-1}}$. This speed is comparable to the circular velocity of the potential, which is $\approx 550 \, {\rm km \, s^{-1}}$ at $r = 30 \, {\rm kpc}$ and $\approx 730 \, {\rm km \, s^{-1}}$ at $r = 60 \, {\rm kpc}$.
\vspace*{1em}
\label{fig-BallisticCondensation}}
\end{figure*}

The origin of this multiphase gas in the power-law zone is not growth of linear thermal instability but rather uplift.  These gas patches originate in the isentropic zone, are uplifted by the outflow, and then follow trajectories that are approximately ballistic.  Black arrows in Figure~\ref{fig-BallisticCondensation} show how one such region rises and falls.  It can be seen rising out of the isentropic zone, reaching $\approx 30$~kpc at $t = 5.10$~Gyr.  Between $t = 5.20$~Gyr and $t = 5.30$~Gyr it reaches an apex at $\approx 80$--$90$~kpc and falls back through $\approx 30$~kpc at $t \approx 5.42$~Gyr.  If the motion of the gas blob were purely ballistic, this excursion in radius would take $\approx 240$~Myr and would require the projectile to rise and fall through $r = 30$~kpc, where the circular velocity is $\approx 550 \, {\rm km \, s^{-1}}$, at a speed $\gtrsim 900 \, {\rm km \, s^{-1}}$.  Instead, the peak speed during both the upward and downward portions of the trajectory is $\sim 500 \, {\rm km \, s^{-1}}$, while most of the time is spent at smaller speeds near the trajectory's turning point.  These deviations from ballistic motion imply that outflow is still providing impetus to the multiphase gas blob during the upward stage, and that drag is limiting the infall speed to $\approx 550 \, {\rm km \, s^{-1}}$ during the downward stage.

It can be argued that these multiphase gas clumps would be deviating more strongly from ballistic trajectories in a more realistic and highly resolved simulation because of hydrodynamic shredding and drag \citep[e.g.][]{Nulsen_1986MNRAS.221..377N,ps05}.  In this simulation, however, drag is clearly not necessary to promote the development of multiphase gas at large altitudes, because the gas is multiphase on both the upward and downward portions of the trajectory.  High-resolution simulations specifically designed to track the hydrodynamical evolution (and perhaps even magnetohydrodynamical evolution) of uplifted gas will be needed to draw more certain conclusions about the roles of uplift and drag in promoting condensation and multiphase structure within the power-law zone. 

\section{Loose Ends}
\label{sec-LooseEnds}

Knowledgeable readers have surely noticed by now that some important physical effects certain to affect the thermal stability of circumgalactic gas are absent from our analysis.  We hope those readers are also wise enough to recognize when a paper has already gotten a little too long.  Very briefly, we will point out some of the missing physics and speculate about its consequences.

\subsection{Angular Momentum}

Rotation of the system, which we have ignored, has at least two major consequences:  (1) locally it can reduce the effective gravitational potential and therefore the stratification on which buoyancy damping depends, and (2) globally it can inhibit accretion of condensed gas onto the central black hole.   Consequence (1) can promote condensation in regions where it otherwise would not happen.  Consequence (2) has the potential to decouple condensation from black-hole fueling.  However, \citet{Gaspari_2015A&A...579A..62G} have explored how rotation affects condensation and chaotic cold accretion using high-resolution simulations of thermally balanced media and show that precipitation can still deliver plenty of cold fuel to the vicinity of the black hole.  The key is for turbulence and cloud-cloud collisions to continually populate the low end of the distribution of specific angular momentum among cold clouds.  This is more easily done when the velocity dispersion of turbulence exceeds the mean rotational speed.  

It is tempting to speculate about the implications of consequence (1) for replenishment of condensed gas in galactic disks.  In a rotating frame of reference, the effective gravitational potential of a spherically symmetric halo with a stellar disk becomes more like that in the plane-parallel simulations of \citet{Meece_2015ApJ...808...43M}.  If the halo contains hydrostatic ambient gas  with $t_{\rm cool}/t_{\rm ff} \gg 1$ then buoyancy damping can inhibit condensation everywhere except the midplane.  Condensation of hot ambient halo gas may then replenish the disk's reservoir of cold gas without any obvious evidence for infall in the form of cool clouds at larger altitudes.  For a more formal analysis of thermal instability in a rotating circumgalactic medium, see \citet{Nipoti_2014ApJ...792...21N}.

\subsection{Inhomogeneous Enrichment}

If cooling of the circumgalactic medium depends strongly on metallicity, as it certainly will in systems with virial temperatures $\lesssim 1$~keV, then inhomogeneities in the distribution of metals will promote thermal instability and condensation of the more highly enriched gas \citep[e.g.,][]{FraternaliBinney_2008MNRAS.386..935F,Marinacci+10,Oppenheimer_2010MNRAS.406.2325O}.    An over-enriched gas blob with the same $K$ and $P$ as the medium surrounding it will cool faster, causing it to sink, and it will not respond as strongly to buoyancy damping.  This coupling between enrichment, entropy, and buoyancy has complex consequences for condensation, meaning that metal transport and mixing in circumgalactic gas will be important to simulate properly in order to clarify the role of inhomogeneous enrichment.

Perhaps counterintuitively, inhomogeneities in enrichment could make it easier for lower-metallicity gas to condense within the galactic disk itself.  This may happen if high-metallicity gas more easily condenses into cold, ballistic clouds at greater altitudes, while lower metallicity gas settles more homogeneously into the disk and condenses there instead.  Again, high-resolution targeted simulations including metal transport will be necessary to resolve this question.

\subsection{Magnetic Fields}

Our entire analysis has implicitly assumed the existence of tangled magnetic fields because we have ignored electron thermal conduction \citep[see][for a discussion of the role of magnetic fields in cooling flows]{Soker10_Bfields}.  In reality, conduction is likely to be somewhat anisotropic without being completely suppressed, which will naturally cause condensation to be filamentary \citep[e.g.,][]{spq10,McCourt+2012MNRAS.419.3319M,Wagh_2014MNRAS.439.2822W}.  Also, larger-scale conduction may help lift cluster cores out of precipitation-regulated states, if the central entropy rises enough for radial heat transport to exceed radiative cooling in the core \citep{Voit+08,gor08,pqs10,Voit_2015Natur.519..203V}.  If that happens, then feedback is no longer necessary to prevent catastrophic cooling.

Another potential consequence of magnetic fields is enhanced drag, which can promote condensation (\S \ref{sec-BobbingBlob}).  This has long been recognized as a possibility \citep[e.g.,][]{Nulsen_1986MNRAS.221..377N}, and recent ALMA observations of remarkably small, sub-Keplerian velocity dispersions among the bulk of the molecular gas clouds in galaxy-cluster cores \citep{McNamara_2014ApJ...785...44M,Russell_2016MNRAS.458.3134R} have rekindled interest in it \citep{McNamara_2016arXiv160404629M}.  Obviously, magnetohydrodynamical simulations are required to clarify whether magnetic tension is strong enough to suspend molecular clouds within a medium that is many orders of magnitude less dense.  

Cosmological MHD simulations will also be needed to clarify whether the magnetothermal instability \citep{Balbus_1991ApJ...372...25B} or heat-flux driven buoyancy instability \citep{Quataert08,pq08} alter the general picture of susceptibility to condensation that we have outlined here.  
When anisotropic conduction is important, buoyancy is more closely related to temperature gradients than to entropy gradients \citep[e.g.,][]{Balbus_2000ApJ...534..420B,Sharma_2009ApJ...699..348S}, which is likely to reduce the dependence of buoyancy damping on $\alpha_K$.  Recent simulations have shown that maximally efficient anisotropic conduction does indeed alter the effects of AGN feedback on the intracluster medium \citep{Kannan_2017ApJ...837L..18K}. 

\subsection{Heat Transport and Dissipation}

We have emphasized the importance of heat transport in stabilizing the CGM of massive galaxies but have not paid any attention to how the transported energy is transformed into heat.  Dissipation of feedback energy in galaxy-cluster cores is currently an active area of research.  Many possibilities are being investigated, including turbulence, shock heating, and mixing of hot gas with cooler gas \citep[e.g.,][]{BanerjeeSharma_2014MNRAS.443..687B,Zhuravleva_2014Natur.515...85Z, YangReynolds_2016arXiv160501725Y, HillelSoker_2016MNRAS.455.2139H}.  These dissipation processes will need to be included in a more complete model.

\newpage

\section{Concluding Thoughts}
\label{sec-Summary}

A summary of the paper's main findings can be found in \S \ref{sec-ReadersGuide}.  Instead of repeating that summary here, we will conclude with some suggestions about how our global model for the ambient circumgalactic medium may be tested with observations and numerical simulations: 
\begin{itemize}

\item {\bf Buoyancy Damping.} The model proposes that buoyancy damping can suppress condensation in circumgalactic media with a significant entropy gradient but allows it to proceed if the median entropy profile is nearly isentropic.  This proposal can be tested with observations of correlations between homogeneity of the ambient medium and the slope of its entropy gradient.  Cluster cores with short central cooling times are ideal for such tests, because both their entropy gradients and their levels of inhomogeneity are observable with X-ray telescopes.  One expects to detect an increasing dispersion in gas entropy and temperature as the large-scale entropy gradient flattens.

\item {\bf Two Precipitation Modes.} Multiphase gas can precipitate out of the circumgalactic medium in two different ways: (1) through growth of thermal instability in the isentropic zone, or (2) through uplift of low-entropy ambient gas.  Isolated clumps of multiphase gas outside of the isentropic zone must therefore originate at lower altitudes and can result from condensation if they are uplifted from regions with $t_{\rm ff} / t_{\rm cool} \lesssim 10$.  Condensation in the power-law zone of gas with greater initial values of $t_{\rm ff} / t_{\rm cool}$ requires either an uplift velocity exceeding $\approx 1.5 v_c$ or some combination of drag and turbulence that slows the infall speed as the uplifted gas falls back toward its original altitude.  These should be the two most prevalent modes of condensation in more complex simulations of galaxy evolution with sufficiently high spatial resolution. 

\item  {\bf Lower Limit on $t_{\rm cool}/t_{\rm ff}$ in the Ambient Medium.} Feedback outbursts tend to produce a lower limit of $\min (t_{\rm cool} / t_{\rm ff}) \approx 10$ in the ambient medium, because ambient gas with lower values of $t_{\rm cool} / t_{\rm ff}$ is vulnerable to uplift-driven condensation.   X-ray observations show that galactic systems with $v_c > 300 \, {\rm km \, s^{-1}}$ generally adhere to this limit.  It should also apply to ambient circumgalactic gas near the virial temperature in lower-mass systems, in which the conditions can be probed by {\em Hubble}-COS observations \citep[e.g.,][]{Stocke_2013ApJ...763..148S,Werk_2014ApJ...792....8W}.

\item {\bf Consequences of Central Thermal Feedback.} Central thermal feedback destabilizes the circumgalactic medium because it expands the isentropic zone and suppresses buoyancy damping.  Condensation can proceed there until feedback raises its cooling time to approximately the current age of the universe.  Simulations implementing this mode of feedback cannot produce realistic galaxy-cluster cores, many of which have central cooling times $\lesssim 1$~Gyr.  Instead, heat deposition by feedback must extend beyond the isentropic zone.  Feedback energy is probably transported there by bipolar outflows, and this is likely to be why kinetic feedback reproduces the characteristics of galaxy cluster cores with much greater success than pure thermal feedback.  Our analysis predicts that any numerical simulation of massive-galaxy evolution in which feedback is centrally injected and purely thermal requires heat input to raise the central cooling time to several Gyr before it succeeds in stopping star formation.  

\item {\bf Quenching and the CGM Entropy Gradient.} Buoyancy damping, when coupled with kinetic feedback, can explain how star formation remains quenched in massive elliptical galaxies with short central cooling times.  If feedback from a bipolar outflow can maintain a significant entropy gradient in the ambient medium, then buoyancy damping will suppress thermal instability and condensation.   Episodic feedback outbursts may occasionally induce condensation through uplift, but the total amount of condensed gas should remain small, as long as the mass of ambient gas in the isentropic zone remains small.  More generally, one expects the time-averaged supply rate of condensed gas from the circumgalactic medium to be similar to the gas mass of the isentropic zone divided by the cooling time of the isentropic zone, as long as that cooling time is significantly less than the age of the universe.  Quenched galaxies with short central cooling times should therefore be observed to have strong entropy gradients and small isentropic cores \citep[as in][]{Werner+2012MNRAS.425.2731W,Werner+2014MNRAS.439.2291W}.

\item {\bf Midplane Condensation in Galactic Disks.} Buoyancy damping may allow the ambient circumgalactic medium to supply a galactic disk with condensing gas via midplane condensation without any obvious ``rainfall.''  This can happen if there is an entropy gradient outside of the midplane and sufficient rotation to inhibit buoyancy damping in the radial direction.  Given enough time, condensation will happen near the midplane (see \S \ref{sec-Midplane}), in a region with a thickness that depends on $( t_{\rm ff} / t_{\rm cool})^2$.  A steady supply of hot ambient gas then settles into the midplane to replenish what is lost to condensation, without any need for infall of cold clouds at large distances from the galactic disk.  This mode of gas supply should be present in numerical simulations of sufficiently high spatial resolution.

\item {\bf Episodic Line Emission from Cooling Gas.} Condensation and star formation in a precipitation-regulated system do not necessarily happen simultaneously, in a nearly steady state (see Figure~\ref{fig-OutburstHistory}).  Line emission from intermediate-temperature ($10^5$-$10^6$~K) condensing gas is expected to be most luminous during the uplift events that trigger rapid condensation.  In between these events, the star-formation rate may exceed the condensation rate for much of the duty cycle, causing a steady decline in the amount of cold gas.

\end{itemize}

\vspace*{0.5em}

We are indebted to 
A. Babul, 
J. Bregman,
A. Crocker, 
A. Evrard, 
M. Fall, 
M. Gaspari, 
A. Kravtsov, 
M. McDonald, 
B. McNamara, 
P. Nulsen, 
P. Oh, 
M. Peeples,
M. Ruszkowski, 
P. Sharma, 
D. Silvia, 
N. Soker,  
M. Sun, 
G. Tremblay, 
J. Tumlinson,
S. White,
and an insightful but anonymous referee
for discussions that have shaped our thinking.
Some of this work was performed at the Aspen Center for Physics, which is supported by National Science Foundation grant PHY-1066293.
GMV, BWO, and MD acknowledge NSF for support through grant AST-0908819 and NASA for support through grants NNX12AC98G, NNX15AP39G, HST-AR-13261.01-A, and HST-AR-14315.001-A.
GLB acknowledges financial support from NSF grants AST-1312888, and NASA grants NNX12AH41G and NNX15AB20G, as well as computational resources from NSF XSEDE, and Columbia University's Yeti cluster.  
Simulations were run on the NASA Pleiades supercomputer through allocation SMD-15-6514 and were performed using the publicly-available Enzo code (http://enzo-project.org) and analyzed with the {\tt yt} package \citep{Turk_yt_2011ApJS..192....9T}. Enzo is the product of a collaborative effort of many independent scientists from numerous institutions around the world. Their commitment to open science has helped make this work possible. 

\appendix

\section{A Toy Model For Reconfiguration of Thermally Balanced Hydrostatic Systems}

In order to gain insight into how cooling and feedback regulate the global configuration of the ambient gas in a galactic gravitational potential, we consider here the behavior of an idealized but representative hydrostatic system that is isentropic within an inflection radius $r_0$ and isothermal outside of that radius.  For a monatomic ideal gas, the equation of hydrostatic equilibrium in spherical symmetry can be written as
\begin{equation}
 \frac {d \ln P} {d \ln r} \: = \: \frac {5} {2} \frac {d \ln T} {d \ln r} - \frac {3} {2} \frac {d \ln K} {d \ln r}
 			\: = \: - 2 \frac {T_\phi} {T} \; \; ,
\end{equation}
where $T_{\phi}$ is defined so that  $2 k T_\phi / \mu m_p = d \phi / d \ln r$.  The temperature gradient is therefore determined by 
\begin{equation}
 \frac {d \ln T} {d \ln r} \: = \: \frac {3} {5}  \alpha_K  - \frac {4} {5} \frac {T_\phi} {T} \; \; ,
\end{equation}
where $\alpha_K \equiv d \ln K / d \ln r$.  Isothermality at temperature $T_0$ can always be enforced in a general potential well by setting $\alpha_K(r) = 4 T_\phi (r) / 3 T_0$.  Instead, we will simplify the mathematics by assuming a potential well with a constant value of $T_\phi$, in which case the temperature of the isothermal region, $T_0 = 4 T_\phi / 3 \alpha_K$, is set by the constant power-law index of the entropy profile.  The system's configuration is then completely specified by the entropy level $K_0$ of the isentropic region, the gas mass $M_0$ within that region, the power-law slope $\alpha_K$ of the entropy profile outside that region, the depth $T_\phi$ of the potential well, and a pressure boundary condition that ensures $T = T_0$ at the outer edge of the isentropic region.

\subsection{The Isentropic Region}

The temperature and electron density profiles in the isentropic region are given by
\begin{equation}
    T(r) = T_0 \left[ 1 - \frac {3 \alpha_K} {5}  \ln \left( \frac {r} {r_0} \right) \right] 
      \; \; \; \; \; , \; \;\; \; \;  
    n_e (r) = \left( \frac {k T_0} {K_0} \right)^{3/2} 
    	\left[ 1 - \frac {3 \alpha_K} {5}  \ln \left( \frac {r} {r_0} \right) \right]^{3/2}
      \; \; ,
\end{equation}
and integrating to obtain the ambient gas mass inside $r_0$ yields
\begin{equation}
  M_0 = \frac {4 \pi} {3}  r_0^3 \, \mu_e m_p \left( \frac {4k T_\phi} {3K_0} \right)^{3/2} 
  			\frac {I_{3/2}(\alpha_K)} {\alpha_K^{3/2}} 
              \; \; \; \; \; , \; \; \; \; \; 
         I_n(\alpha_K) \equiv 
             \int_0^1 \left( 1 - \frac {3 \alpha_K} {5}  \ln x \right)^{n} 3x^2 dx  
         \; \; ,
  \label{eq-M0}
\end{equation}
with $I_{3/2}(\alpha_K) \approx (1 + 0.33 \, \alpha_K^{1.1})$ to within 0.2\% in the range $2/3 < \alpha_K < 2$.   Reorganizing the equation for $M_0$ then gives $r_0$ in terms of the three adjustable parameters ($K_0$, $M_0$, $\alpha_K$) that control the system's configuration.  

In most astrophysical systems of interest, the $t_{\rm cool} / t_{\rm ff}$ ratio declines with radius in isentropic gas.  In this particular system, the dependence of cooling time on radius in the isentropic region follows
\begin{equation}
  \frac {d \ln t_{\rm cool}} {d \ln r}   
    \; = \; 
   (1 + 2 \lambda) \frac {3 \alpha_K} {10}   
      \left[ 1 - \frac {3 \alpha_K} {5}  \ln \left( \frac {r} {r_0} \right) \right]^{-1}
    \; \; ,
\end{equation} 
where $\lambda \equiv d \ln \Lambda / d \ln T$.  The condition for $t_{\rm cool} / t_{\rm ff}$ to decline with radius in a system with constant $T_\phi$ is $d \ln t_{\rm cool} / d \ln r < 1$.  A decline with radius is therefore guaranteed for $\alpha_K < 5/3$, since astrophysical cooling functions have $\lambda \leq 1/2$ at the temperatures characteristic of ambient circumgalactic gas. 

Buoyancy damping does not suppress thermal instability in this part of the system.  It is therefore prone to producing multiphase gas via condensation on a timescale $\sim \ln (1 /\delta_K) \times t_{\rm cool}$, where $\delta_K$ is the fractional amplitude of the initial entropy fluctuations.

\subsection{The Isothermal Region}

The structure of the system's isothermal region outside of $r_0$ is given by 
\begin{equation}
  K(r) = K_0 \left( \frac {r} {r_0} \right)^{\alpha_K} 
  		\; \; \; \; \; , \; \;\; \; \;  
  n_e(r) = \left( \frac {k T_0} {K_0} \right)^{3/2} \left( \frac {r} {r_0} \right)^{-3 \alpha_K /2} 
                   \; \; \; \; \; , \; \; \; \; \; 
  \frac {t_{\rm cool}} {t_{\rm ff}} \propto r^{(3 \alpha_K - 2)/2}   
                                          \; \; \; . 
  \label{eq-IsothermalRegion}
\end{equation}
Buoyancy damping suppresses condensation in the isothermal region as long as $\alpha_K \gtrsim (t_{\rm ff} / t_{\rm cool})^2$.  Suppression of condensation therefore applies to the entire isothermal region if $\alpha_K > 2/3$ and $t_{\rm cool} / t_{\rm ff} \gg 1$ at $r_0$. In such systems, the minimum value of $t_{\rm cool} / t_{\rm ff}$ is at $r_0$.   


\subsection{Redistribution of Heat}

In order for the system to remain in overall thermal balance as it evolves, the heat energy lost by one of its subsystems must be gained by another subsystem.  Let subsystem 1 with gas mass $\delta M_g$ be the one that loses an amount of heat energy $\delta Q$ and experiences a corresponding change in specific entropy from $K_1$ to $K_1 - \delta K_1$, such that
\begin{equation}
  \delta Q = \frac {3k} {2}  \frac {\delta M_g} {\mu_e m_p} 
  				\int_{K_1-\delta K_1}^{K_1} T_1(K) \frac {d K} {K} \; \; .
\end{equation}
In general, heat transfer changes the temperature $T_1$ of the subsystem, and the relationship between $\delta Q$ and $\delta K$ depends on $T_1(K)$ during the heat-transfer episode.  For example, the particular case of condensation at constant pressure corresponds to $T_1 \propto K^\eta$ with $\eta = 3/5$, and condensation at constant volume corresponds to $\eta = 1$.   
In either case, a condensation process in which $K_1 - \delta K_1 \ll K_1$ leads to
\begin{equation}
  \delta Q = \frac {3} {2}  \frac {kT_1(K)} {\eta} \frac {\delta M_g} {\mu_e m_p} 
  				 \; \; . 
\end{equation}
A more complex condensation process can be characterized by the effective value of $\eta$ for which this equation holds.

Let subsystem 2 be that one that gains heat energy $\delta Q$.  If subsystem 2 has a gas mass $M_g \gg \delta M_g$, then the gain in heat energy raises its specific entropy by
\begin{equation}
  \delta K_2 =  \frac {K_2} {\eta}  \frac {T_1(K_1)} {T_2(K_2)} \frac {\delta M_g} {M_g} 
  				 \; \; . 
\end{equation}
Notice that this equation can also be used to derive the relation $M_g \propto K^{-\eta}$ between the specific entropy $K$ and gas mass $M_g$ of the ambient medium in a system that is kept in global thermal balance as condensation gradually removes mass from the ambient phase and adds it to the condensed phase.

\subsection{Reconfiguration of Thermally Balanced Equipotential Layers}

Heat redistribution algorithms that keep each equipotential layer in independent thermal balance \citep[e.g.,][]{McCourt+2012MNRAS.419.3319M,Sharma_2012MNRAS.420.3174S,Gaspari+2013MNRAS.432.3401G,Meece_2015ApJ...808...43M} allow gas to condense out of the ambient phase in the isentropic region and produce an accompanying rise in $K_0$.  Meanwhile, the specific entropy $K$ of each Lagrangian layer outside of the isentropic region remains unchanged because buoyancy damping causes entropy fluctuations there to saturate at a fractional amplitude $\delta_K \sim (t_{\rm ff} / t_{\rm cool})^2$ instead of condensing.  Consequently, a rise in $K_0$ drives convective mixing in the neighborhood of $r_0$ as gas with $K < K_0$ sinks into the isentropic zone and adds to $M_0$.

Those changes in system configuration are best analyzed in the gas-mass domain.  Let $M_g(K)$ represent the mass of ambient gas with specific entropy less than $K$, and let $M_c$ be the mass of gas that has already condensed. Inverting $M_g(K)$ gives a two-segment entropy distribution which is isentropic at $K_0$ for $M_g < M_0$ and follows a power law with $d \ln K / d \ln M_g = 2 \alpha_K / (6-3 \alpha_K)$ for $M_g > M_0$.  A condensation event that raises $K_0$ by $\delta K_0$ adds gas mass $\delta M_c = \eta M_0 \, \delta K_0 / K_0$ to the condensed phase and changes $M_0$ by
\begin{equation}
   \delta M_0 \; =  \; M_0 
   	\left( \frac {6-3\alpha_K} {2 \alpha_K}  - \eta \right) 
	\frac {\delta K_0} {K_0}  \; \; .
\end{equation}
Condensation therefore drives the ratio of condensed to isentropic gas toward $M_c / M_0 = 2 \alpha_K \eta / (6 - 3 \alpha_K)$, implying that the condensed gas mass is comparable to $M_0$ for representative values of $\alpha_K \sim 1$ and $\eta \approx 3/5$.

As $K_0$ rises, there is a particular value of $\eta$ which ensures that the $K(r)$ profile remains isothermal,   
\begin{equation}
  \eta_0 \; = \; \frac {3} {2} - \frac {3} {\alpha_K} \left[ 1 - \frac {1} {I_{3/2}(\alpha_K)} \right]
  	\; \approx \; \frac {3} {2} \left[
	         \frac {3 + \alpha_K^{1.1}  - 2 \alpha_K^{0.1}} {3 + \alpha_K^{1.1}} 
	         \right]  \; \; ,
\end{equation}
because it ensures that $M_0 + M_c = M_0 [2/(2-\alpha_K)] I_{3/2}^{-1}(\alpha_K)$, which is the gas mass that would be contained within $r$ if the isothermal region extended all the way to $r = 0$.
For $\alpha_K \approx 1$, this equation gives $\eta_0 \approx 3/4$, which is in between the characteristic values for constant-pressure condensation ($\eta = 3/5$) and constant-volume condensation ($\eta = 1$). Evolution of $K(r)$ in a system with thermally balanced equipotential layers therefore remains close to the idealized isentropic/isothermal configuration, with $K(r)$ staying nearly fixed at $r > r_0$ as condensation in lower-lying layers raises the value of $K_0$ \citep[see, e.g.,][]{Sharma_2012MNRAS.420.3174S,Gaspari+2013MNRAS.432.3401G}.  Deviations of the actual value of $\eta$ from $\eta_0$ cause $M_0$ to deviate somewhat from the required value for maintaining the idealized configuration and result in small deviations from isothermality just outside of $r_0$.

The rate at which such a system's $K(r)$ profile evolves depends primarily on $t_{\rm cool} (K_0)$ but also on the initial sizes of the entropy fluctuations that lead to condensation.  If the fractional amplitudes of these fluctuations are of order unity, then $d \ln K_0 / dt \approx t_{\rm cool}^{-1} (K_0)$. If instead the seed fluctuations are smaller, then more time is required for condensation.  The results of \S 4 show that buoyancy damping leads to saturation of perturbation growth at smaller amplitudes for larger values of $t_{\rm cool} / t_{\rm ff}$.  Condensation at the center of the system, where the entropy gradient vanishes, is still possible but requires a time period that grows progressively longer compared to the cooling time as $t_{\rm cool} / t_{\rm ff}$ increases.

\subsection{Reconfiguration of a Globally Balanced System}

A globally balanced system in which one layer gains heat energy while another layer loses the same amount of heat energy has more freedom to change configuration.  Qualitatively, the two simplest kinds of configuration changes are: (1) if cooling exceeds heating inside of $r_0$, causing $K_0$ to decrease, then heating must exceed cooling in the outer region, or (2) if heating exceeds cooling inside of $r_0$, causing $K_0$ to increase, then cooling must exceed heating in the outer region.  

In our toy model, a small change in entropy $\delta K(M_g)$ as a function of the gas mass $M_g$ enclosed within a given radius corresponds to a change in heat energy given by
\begin{equation}
  \delta Q \; = \;  \frac {3} {2} \frac {k T_0} {\mu m_p} 
                            \left[ \delta \ln K_0  \cdot \int_0^{M_0} \frac {T} {T_0} \, d M_g 
                            \: + \:
                            \int_{M_0}^{M_{\rm b}} 
                              \delta \ln K \cdot d M_g   \right]
                           \; \; , 
\end{equation}
where $M_{\rm b}$ is the ambient gas mass within the boundary radius $r_{\rm b}$.
The configuration changes permitted by global thermal balance within radius $r_{\rm b}$ have the property
\begin{equation}
      \int_{M_0}^{M_{\rm b}}  \delta \ln K (M_g) \cdot d M_g   
                                   \: = \:
           -   M_0 \, \frac {I_{5/2}(\alpha_K)} {I_{3/2}(\alpha_K)}   \, \delta \ln K_0
                           \; \; ,
     \label{eq-GlobalBalance}
\end{equation}
where $I_{5/2}(\alpha_K)$ and $I_{3/2}(\alpha_K)$ are both given by the definition in equation (\ref{eq-M0}).   From equations (\ref{eq-M0}) and (\ref{eq-IsothermalRegion}) we obtain an expression for the entropy-profile change in terms of changes in the three model parameters:
\begin{equation}
  \delta \ln K = \left( 1 - \frac {\alpha_K} {2} \right) \delta \ln K_0
  			- \frac {\alpha_K} {3} \, \delta \ln M_0
			- \left[ \ln \left( \frac {K} {K_0} \right) - \frac {\alpha_K} {2} 
			    + \frac {\alpha_K} {3} \frac {d \ln I_{3/2}} {d \ln \alpha_K} \right] \delta \ln \alpha_K
	\; \; .
  \label{eq-DeltaLnK}
\end{equation}
Integrating this function over $M_g$ then gives
\begin{equation}
  \left( \frac {M_b - M_0} {M_0} \right) \, \delta \ln K_b  
     -  \left( \frac {2 \alpha_K} {6 - 3 \alpha_K} \right) 
             \left[ \left( \frac {M_b - M_0} {M_0} \right) + \frac {3} {\alpha_K I_{3/2}} \right]  \delta \ln \alpha_K
     \; = \; - \frac {I_{5/2}} {I_{3/2}} \delta \ln K_0
 \; \; ,
   \label{eq-GlobalBalanceEquation}
\end{equation}
where $\delta \ln K_b$ is the change in specific entropy at the boundary, obtained by substituting $K_b \equiv K(r_{\rm b})$ into equation~({\ref{eq-DeltaLnK}).
Parameter adjustments satisfying this equation change the system's configuration without changing the amount of heat energy it contains.

Some configuration changes require a flow of free energy through the system, while others do not.  The change in the system's global entropy $S$ when $K_0$ changes under conditions of global thermal balance is
\begin{equation}
  \delta S \; = \; \frac {3 k} {2 \mu m_p} \int_0^{M_{\rm b}} \frac {\delta K} {K} \: d M_g 
               \; = \; - \frac {3 k M_0} {2 \mu m_p} 
               		\left[ \frac {I_{5/2}} {I_{3/2}} - 1 \right]
			\delta \ln K_0 \; \; .  
\end{equation}
Reduction of $K_0$ can proceed without a source of free energy (since $I_{5/2} > I_{3/2}$) because it corresponds to a transfer of heat from the higher temperature isentropic zone to the lower temperature isothermal zone, thereby boosting the global entropy.  However, growth of $K_0$ requires a source of free energy, because the lower-temperature outer regions then lose heat while the higher-temperature inner regions gain it, resulting in $\delta S < 0$.

With the global balance constraint in place, the system has only two degrees of freedom.  Applying another constraint at the boundary leaves only one degree of freedom.  One option is to constrain 
the temperature at the outer boundary to remain at $T_0$, which has the consequence of holding $\alpha_K$ fixed.  In that case, equation~(\ref{eq-GlobalBalance}) reduces to
\begin{equation}
  \delta \ln M_0 =  \left[ \left(\frac {6 - 3 \alpha_K} {2 \alpha_K} \right)
  				+ \frac {3} {\alpha_K} \frac {I_{5/2}} {I_{3/2}}
				   \left( \frac {M_0} {M_b - M_0} \right) \right] 
			    \delta \ln K_0 \; \; , 
\end{equation}
and central heating with $\alpha_K < 2$ causes both $K_0$ and $M_0$ to rise while $\alpha_K$ stays fixed.  Alternatively, holding the entropy constant at the boundary gives
\begin{eqnarray}
  \delta \ln K_0 & \; = \; & \left( \frac {2} {2- \alpha_K} \right)
                                          \left[ \frac {I_{3/2}} {I_{5/2}} \left( \frac {M_b - M_0} {M_0} \right)
						+ \frac {3} {I_{5/2} \alpha_K} \right] 
					\, 
					\delta \ln \alpha_K
					\\
  \delta \ln M_0 & \; = \; & \frac {3} {\alpha_K}
  				   \left[ \frac {I_{3/2}} {I_{5/2}} 
				             \left( \frac {M_b - M_0} {M_0} \right) + \frac {3} {I_{5/2} \alpha_K}
				      -  \ln \frac {K_b} {K_0} + \frac {\alpha_K} {2} 
			               - \frac {\alpha_K} {3} \frac {d \ln I_{3/2}} {d \ln \alpha_K} \right] \, 
				  \delta \ln \alpha_K \; \; .
\end{eqnarray}
For this boundary condition, $M_0$ must again increase as $K_0$ grows in order for the system to remain in global thermal balance.  The entropy slope $\alpha_K$ must increase as well, but that change is small if $M_b \gg M_0$.

This result has important qualitative implications for feedback-regulated systems.  If they are thermally balanced by a central heat source, then central heating makes ambient gas there isentropic, and the values of both $K_0$ and $M_0$ increase with time.  Condensation in that region is not suppressed by buoyancy damping and can add to the mass $M_c$ of condensed gas at a rate $\dot{M}_c \sim M_0 / t_{\rm cool} (K_0)$ until the rise in $K_0$ is sufficient for $t_{\rm cool} (K_0)$ to approach a Hubble time.  This is not a finely regulated feedback loop but rather a monotonically evolving system in which feedback gradually shuts off condensation while a potentially large amount of condensed gas accumulates.

In contrast, globally balanced systems in which heating exceeds cooling outside of the isentropic zone generally have values of $K_0$ and $M_0$ that {\em decrease} with time.  Buoyancy damping limits condensation to a relatively small central region, and the associated cooling time also decreases with time, with $t_{\rm cool} \stackrel {\propto} {\sim} K_0^{3/2}$.  Ultimately, something must prevent the central cooling time in the ambient medium from becoming singular.  One lower limit on the cooling time of ambient gas at small radii is $t_{\rm cool} \gtrsim t_{\rm ff}$, because otherwise thermal instability is unavoidable.  Another is stochastic heating that occasionally exceeds cooling, even if cooling exceeds heating during most of the duty cycle.  In a less idealized system, the ambient cooling time that prevails after a heating outburst will be the timescale on which coupling and self-regulation happen, and this timescale can be much less than a Hubble time. 

\subsection{Global Regulation}

Stable global regulation at a particular $K_0$ and $r_0$ requires the heat input balancing radiative cooling to be divided between the isentropic and isothermal zones so that each zone is individually balanced. The radiative luminosity $L$ of each zone can be calculated by integrating
\begin{equation}
  \frac {dL} {d \ln r} = 4 \pi \left( \frac {n_i} {n_e} \right) \left( \frac {r} {K} \right)^3 (kT)^3 \Lambda (T)
\end{equation}
over radius.  This results in a luminosity
\begin{equation}
  L_0 = \frac {4 \pi} {3}  \left( \frac {n_i} {n_e} \right) 
              \left( \frac {r_0} {K_0} \right)^3 (kT_0)^3 \Lambda (T_0)
  		\, I_{3+\lambda}(\alpha_K)  
\end{equation}
for the isentropic zone and a total luminosity
\begin{equation}
  L = L_0 \left[ 1 + \frac {3 \ln (K_b / K_0)} {I_{3+\lambda}(1)} \right]
\end{equation}
for the particular case of $\alpha_K = 1$.  In order to remain in a configuration with a particular value of $K_0$, the fraction $f_0$ of the globally balancing heat input that is deposited into the isentropic zone must be equal to $L_0/L$.

Specifying a function $f_0(K_0)$ to represent how the feedback required for global balance distributes heat as a function of radius places one more constraint on the system and leads to a potentially unique solution for the global configuration. However, not all functions $f_0(K_0)$ permit a stable solution.  Global stability requires a heat-distribution response with the property
\begin{equation}
  \frac {d \ln f_0} {d \ln K_0} < \frac {3 f_0} {I_{3+\lambda} (1)}
\end{equation}
so that upward perturbations of $K_0$ lead to net cooling of the isentropic zone and downward perturbations lead to net heating.  If that happens, then the system's global configuration remains stable, with a value of $K_0$ satisfying
\begin{equation}
  f_0(K_0) \approx \left[ 1 + \frac {3 \ln ( K_b / K_0)} {I_{3+\lambda}(1)} \right]^{-1}  \; \; .
\end{equation}
Qualitatively, the requirement for global stability corresponds to a heat distribution mechanism in which heat deposition into the isentropic zone does not rise very much, and perhaps even declines, as $K_0$ rises.  

In the context of a configuration with $\alpha_K \approx 1$, this feature is a physically plausible consequence of feedback energy transport with a bipolar outflow, because the column density of the isentropic zone is $N_0 \propto r_0 K_0^{-3/2} \propto K_0^{(2 - 3 \alpha_K)/(2 \alpha_K)}$.  For systems with $\alpha_K > 2/3$, a rise in $K_0$ therefore causes $N_0$ to decrease.  The isentropic zone then becomes progressively less able to capture and thermalize the kinetic energy of the outflow as $K_0$ rises, which enables global regulation.

\end{document}